\documentclass[fleqn,usenatbib]{mnras}

\usepackage{newtxtext,newtxmath}
 
\usepackage[T1]{fontenc}
\usepackage{ae,aecompl}
\usepackage{subfig}

\DeclareRobustCommand{\VAN}[3]{#2}
\let\VANthebibliography\thebibliography
\def\thebibliography{\DeclareRobustCommand{\VAN}[3]{##3}\VANthebibliography}
\providecommand{\abs}[1]{\lvert#1\rvert}

\usepackage{graphicx}	
\usepackage{amsmath}	
\usepackage{amssymb}	

\title[Coorbital thermal torques: a numerical study]{Numerical study
  of coorbital thermal torques on cold or hot satellites}

\author[R. O. Chametla and F. S. Masset]{
Ra\'ul O. Chametla,$^{1}$\thanks{E-mail: rortegaesfm@gmail.com (ROC)}
Fr\'ed\'eric S. Masset,$^{1}$
\\
$^{1}$Instituto de Ciencias F\'isicas, Universidad Nacional Aut\'onoma de M\'exico, Av. Universidad s/n, 62210 Cuernavaca, Mor., M\'exico
}

\date{Accepted XXX. Received YYY; in original form ZZZ}

\pubyear{2020}

\begin{document}
\label{firstpage}
\pagerange{\pageref{firstpage}--\pageref{lastpage}}
\maketitle

\begin{abstract}
  We evaluate the thermal torques exerted on low-mass planets embedded
  in gaseous protoplanetary discs with thermal diffusion, by means of
  high-resolution three-dimensional hydrodynamics simulations. We
  confirm that thermal torques essentially depend on the offset
  between the planet and its corotation, and find a good agreement
  with analytic estimates when this offset is small compared to the
  size of the thermal disturbance. For larger offsets that may be
  attained in discs with a large pressure gradient or a small thermal
  diffusivity, thermal torques tend toward an asymptotic value broadly
  compatible with results from a dynamical friction calculation in an
  unsheared medium. We perform a convergence study and find that the
  thermal disturbance must be resolved over typically $10$~zones for a
  decent agreement with analytic predictions.  We find that the
  luminosity at which the net thermal torque changes sign matches that
  predicted by linear theory within a few percents. Our study confirms
  that thermal torques usually supersede Lindblad and corotation
  torques by almost an order of magnitude for low mass planets. As we
  increase the planetary mass, we find that the ratio of thermal
  torques to Lindblad and corotation torques is progressively reduced,
  and that the thermal disturbance is increasingly distorted by the
  horseshoe flow. Overall, we find that thermal torques are dominant
  for masses up to an order of magnitude larger than implemented in
  recent models of planetary population synthesis. We finally briefly
  discuss the case of stellar or intermediate-mass objects embedded in
  discs around AGNs.
\end{abstract}




\section{Introduction}
Protoplanets gravitationally interact with the protoplanetary discs
out of which they form. This interaction leads to a progressive change
of their orbital elements, in particular of their semi-major
axis. This process, known as planetary migration, has been extensively
studied since the realisation by \citet{gt80} that the semi-major axis
could vary by large factors over the lifetime of the protoplanetary
disc. Most analytic studies focused on planets embedded in isothermal
or adiabatic discs (the disc's torque is essentially the same in these
two limit cases, to within a factor $\gamma$, the adiabatic index). It
was found that for realistic disc profiles, the disc's torque is
generally a negative quantity that leads to a decrease of the planet's
orbital radius with time \citet{ww86,tanaka2002}. Protoplanetary discs
are neither isothermal nor adiabatic, however. They are subjected to
heat transport, essentially effected by radiative transfer. Thermal
diffusion depends on a number of parameters, such as the disc's
temperature and density, and the opacity of its dust component.
Planet forming regions of protoplanetary discs are thought to have a
thermal diffusivity of same order, or larger than their kinematic
viscosity \citet{pbk11}. While early attempts to include thermal
diffusion in theories of planet migration were limited to its impact
on the non-linear corotation torque exerted on intermediate-mass
planets \citep{2010ApJ...723.1393M,pbk11,2017MNRAS.471.4917J}, recent
studies of the interaction between a low-mass planet and a disc with
thermal diffusion have shown that the torque exerted on the planet
differs significantly from the torque that would be exerted if the
disc was adiabatic \citep{2014MNRAS.440..683L,
  2017MNRAS.472.4204M}. If, furthermore, the planet releases energy
into the surrounding gas, another component of the disc's force onto
the planet appears, that scales with its luminosity. If the planet is
on a circular orbit, this force leads to a torque, called heating
torque, that is in general positive. Low-mass planets undergoing
planetesimal or pebble accretion (or both) at rates of order
$10^{-5}\;\mathrm{M}_\oplus.\mathrm{yr}^{-1}$ can have luminosities
large enough to reverse the net torque and undergo outward migration
\citep{2015Natur.520...63B,2017MNRAS.472.4204M}. If the planet is free
to move on a non-circular, non-coplanar orbit, this force can lead to
a growth of the planet's eccentricity
\citep{2017MNRAS.465.3175M,2017A&A...606A.114C,2017arXiv170401931E,2019MNRAS.485.5035F}
and inclination
\citep{2017MNRAS.465.3175M,2017arXiv170401931E,2019MNRAS.485.5035F}.

In the case of a non-luminous planet embedded in a disc with thermal
diffusion, the torque difference with respect to the adiabatic case is
called the cold thermal torque. Compared to the adiabatic case, the
gas in the vicinity of the planet is colder and denser, as the energy
arising from compressional heating diffuses away from the planet.  The
cold thermal torque arises from the density difference between that of
the flow with thermal diffusion and that of the adiabatic flow. This
density difference displays a striking similarity with the
perturbation of density arising from the release of heat by the
planet, except for the sign: the latter corresponds to a negative
perturbation of density, centred on the planet, and exhibiting two
lobes in the downstream parts of the Keplerian flow, whereas the
former has same shape, but has a positive sign.

When the planet is offset from corotation as is the case when there is
a radial pressure gradient, the lobes are not symmetric with respect
to the planet. The lobe
located on the same side of corotation as the planet is more
pronounced, so that there is a net torque on the planet with same sign
as that arising from this lobe. In the usual case in which the disc is
slightly sub-Keplerian and the planet is outside of its corotation,
the heating torque is positive \citep{2015Natur.520...63B}, while the
cold thermal torque is negative \citep{2014MNRAS.440..683L}.

\citet{2017MNRAS.472.4204M} provides analytic expressions for both
torques, using linear perturbation theory. These expressions require a
hierarchy of length scales: the distance $x_p$ of the planet to
corotation must be much smaller than the size $\lambda_c$ of the
thermal disturbance, which itself must be much smaller than the
pressure lengthscale $H$ of the disc. The purpose of the present work
is to corroborate numerically the analytic expressions, to assess the
resolution required to correctly capture thermal torques, and to
identify possible deviations of the thermal torque values with that
predicted by linear theory. In a somehow similar spirit,
\citet{2019MNRAS.483.4383V,2020arXiv200413422V} have realised
numerical simulations to corroborate the heating force exerted on a
perturber moving in a three-dimensional medium at rest
\citep{2017MNRAS.465.3175M}. This process is more easily described
than that occurring in a protoplanetary disc, due to the lack of shear
and to the existence of axial symmetry. Here, we specifically focus on
the more challenging problem of the release of heat in the sheared
flows of differentially rotating protoplanetary discs.

In a recent work,
\citet{2020arXiv200503785H} investigated the heating torque by means
of numerical simulations. As they focused exclusively on the heating
component, they considered a massless, point-like heat source in a
shearing box. Here we adopt a different approach: we consider the full
problem of a low (but non vanishing) mass planet embedded in a disc
with thermal diffusion. This enables us to study not only the heating
torque but also the cold thermal torque. While we do not consider a
full disc, for reasons of computational cost, we consider a wedge of a
spherical mesh centred on the planet, rather than a shearing box.

Our paper is organised as follows. In
Section~\ref{sec:initial-conditions}, we present our setup.
In section~\ref{sec:summary-results-from} we summarise the results
on thermal torques obtained by linear theory.
In
Section~\ref{sec:heating-torque}, we perform a study of the heating
torque and compare our results to analytic expectations.  In
Section~\ref{sec:cold-thermal-torque}, we perform a study of the total
torque, which uncovers the cold thermal torque.
In section~\ref{sec:toward-larger-masses} we examine the case of
larger mass planets, then we discuss in
section~\ref{sec:discussion} the importance of thermal torques
in two different astrophysical contexts: that of low mass protoplanets
embedded in protoplanetary discs, and that of stellar mass objects
embedded in AGN discs, and we summarise our results in
section~\ref{sec:conclusions}.

\section{Problem setup}
\label{sec:initial-conditions}
We present here the different parts of our setup: the disc, the
planet, and we present the code used to solve the equations of hydrodynamics.
\subsection{Protoplanetary disc}
\label{sec:protoplanetary-disc}
We consider a three-dimensional, non self-gravitating inviscid gaseous disc
whose motion is governed by the following equations: the equation of
continuity which reads
\begin{equation}
    \partial_t\rho+\nabla\cdot(\rho \mathbf{v})=0,
	\label{eq:continuity}
\end{equation}
the equation of conservation of momentum which reads
\begin{equation}
    \partial_t(\rho\mathbf{v})+\nabla\cdot(\rho\mathbf{v}\otimes\mathbf{v}+p\mathsf{I})=-\nabla p -\rho\nabla\Phi,
	\label{eq:momentum}
\end{equation}
and the equation of evolution of the  density of internal energy $e$,
which reads:
\begin{equation}
    \partial_te+\nabla\cdot(e \mathbf{v})=-p\nabla\cdot\mathbf{v}-\nabla\cdot\mathbf{F}_H+\mathbf{S}.
	\label{eq:energy}
\end{equation}
In these equations $\rho$, $\mathbf{v}$ and $\Phi$ denote the density, the velocity and the gravitational
potential, respectively. The source term for the energy (arising from the heat
release of the planet) is denoted with $\mathbf{S}$ and $\mathsf{I}$ represents the unit tensor. Here $\mathbf{F}_H$ is the heat flux,
given by:
\begin{equation}
    \mathbf{F}_H=-\chi\rho\nabla\left(\frac{e}{\rho}\right),
	\label{eq:flux}
\end{equation}
where $\chi$ is the thermal diffusivity. In the planet forming regions
of protoplanetary discs, the heat flux usually arises from radiative
diffusion in the optically thick regions between the disc's
photospheres. In that case the thermal diffusivity depends on the
temperature, density and opacity of the gas
\citep[e.g.][Eq.~34]{2017MNRAS.471.4917J}. Here we adopt a constant
thermal diffusivity, which allows a clean comparison with analytic
results. This approximation is valid for weakly perturbed parts of the
flow, i.e. essentially outside of the planetary Bondi radius. It is
therefore justified in the present work, as most of the planets
considered here have a Bondi radius much smaller than the thermal
disturbance.

The gas pressure $p$ obeys the equation of state
of ideal gases:
\begin{equation}
p=(\gamma-1)e,
 \label{eq:pressure}
\end{equation}
where $\gamma$ is the adiabatic index. Unless otherwise stated we use
in the following $\gamma=7/5$, appropriate for the diatomic molecules
that constitute most of the gaseous disc. We neglect the radiation
pressure, which is negligible in the cold, dense environment of
protoplanetary discs.

We use spherical coordinates $(r,\theta,\phi)$, where $r$ is the
radial distance from the star, $\theta$ is the polar angle
($\theta=\frac{\pi}{2}$ at the midplane of the disc) and $\phi$ is the
azimuthal angle. We assume different key quantities of the disc to be
power laws of the radial distance $r$.
The aspect ratio $h \equiv H/r$, where $H$ is the vertical scale height
of the disc and $r$ the distance to the central star, obeys the power law:
\begin{equation}
  \label{eq:1}
  h=h_p\left(\frac{r}{r_p}\right)^f,
\end{equation}
where $f$ is the flaring index, $r_p$ is the orbital radius of the
planet and $h_p$ is the aspect ratio at the planet's location. In all
the simulations presented here, we have $h_p=0.05$. Eq.~\eqref{eq:1}
implies that  the temperature is also a power law of the distance to
the star:
\begin{equation} 
T(r)=T_0\left(\frac{r}{r_p}\right)^{-\beta},  
 \label{eq:Temperature}
\end{equation}
  where $T_0$ is the temperature at the planet's location and the
  exponent $\beta$ is related to the flaring index through:
  \begin{equation}
    \label{eq:2}
    \beta=1-2f.
  \end{equation}
  Similarly, the surface density is also chosen to be a power law of
  $r$:
  \begin{equation}
    \Sigma(r)=\Sigma_0\left(\frac{r}{r_p}\right)^{-\alpha},
    \label{eq:Sigma}
  \end{equation}
  where $\Sigma_0$ is the surface density at $r=r_p$. 

\subsection{Planet and stellar potentials}
\label{sec:planet}

The gravitational potential $\Phi$ due to the central star and
the planet is given by
\begin{equation}
\Phi=\Phi_S+\Phi_p,
 \label{eq:potential}
\end{equation}
where
\begin{equation}
\Phi_S=-\frac{GM_\star}{r},
 \label{eq:Star_potential}
\end{equation}
$M_\star$ being the mass of the central star and $G$ the gravitational constant,
and
\begin{equation}
\Phi_p=-\frac{GM_p}{\sqrt{r'^2+\epsilon^2}}+\frac{GM_pr\cos\phi\sin\theta}{r_p^2}
 \label{eq:Planet_potential}
\end{equation}
are respectively the stellar and planetary potential. In
Eq.~\eqref{eq:Planet_potential}, $M_p$ is the planet mass,
$r'\equiv\abs{\mathbf{r}-\mathbf{r}_p}$ is the distance to the planet,
$\phi$ is the azimuth with respect to the planet's direction, $\theta$
is the colatitude and $\epsilon$ is a softening length used to avoid
computational problems arising from a divergence of the potential in
the vicinity of the planet. The second term on the right hand side of
Eq.~\eqref{eq:Planet_potential} is the indirect term arising from the
reflex motion of the star.  For the lowest mass planets considered in
this work, we have performed simulations with two different values of
the softening length: $\epsilon_1 = 0.1H$ and
$\epsilon_2=5\cdot 10^{-3}H$. The former is comparable to the size of
the radial disturbance, while the latter is comparable to the mesh
resolution, and is typically an order of magnitude smaller than the
size of the thermal disturbance. In all cases the back torque exerted
on the planet by the perturbed disc was evaluated using the softening
length $\epsilon_2$. We have found that thermal torques thus evaluated
do not depend on the softening length, and that there are only minute
differences in the perturbation of the density field arising from heat
release. All runs with larger mass planets as well as those focused on
the study of the cold torque were performed using a softening length
$\epsilon_2$.

\subsection{Code and mesh domain}
\label{sec:code-mesh-domain}

To numerically solve equations
(\ref{eq:continuity}-\ref{eq:pressure}), we use the hydrodynamic
public code {FARGO3D\footnote{http://fargo.in2p3.fr}}
\citep{2016ApJS..223...11B} with orbital advection enabled
\citep{fargo2000}. We have implemented the energy diffusion equation
as an additional source step corresponding to the following
differential equation:
\begin{equation}
\partial_t e=\nabla\cdot\mathbf{F}_H
 \label{eq:diffusion}
\end{equation}
where $\mathbf{F}_H$ is given by equation (\ref{eq:flux}).

In our simulations we adopt parameters for the pressure lengthscale
and for the thermal diffusivity that are typical of those found at a
few astronomical units in protoplanetary discs
\citep[e.g.][]{2015A&A...575A..28B,2015MNRAS.452.1717L,2015Natur.520...63B}.
As a consequence the size of the thermal disturbance $\lambda_c$ is
significantly smaller than the pressure lengthscale $H$
\citep{2017MNRAS.472.4204M}. This is in contrast with the work of
\citet{2020arXiv200503785H}, who considered a thermal disturbance
comparable in size to the pressure lengthscale, which allowed them to
satisfy largely the scale separation between the distance $x_p$
of the planet to corotation, and $\lambda_c$,
i.e. they have $x_p\ll \lambda_c \lesssim H$, whereas in our case we
have $x_p\lesssim \lambda_c \ll H$. Namely, we take the following
values in all our runs:
\begin{equation}
  \label{eq:3}
  \chi = 10^{-5}r_p^2\Omega_p,
\end{equation}
where $\Omega_p$ is the planet's frequency.
At the planet's location we have:
\begin{equation}
  \label{eq:4}
  \lambda_c = 2.2\cdot 10^{-3}r_p\approx\frac{H}{23}.
\end{equation}
Resolving this lengthscale over a few zones while simulating a full
disc would have a prohibitive computational cost, especially for
parameter space explorations such as those presented here. For this
reason we restrict our computational domain in the three
coordinates. In the $r$-direction we cover the interval
$[r_\mathrm{min},r_\mathrm{max}]=[0.9r_p,1.1r_p]$, in the $\theta$-direction we use
$[\theta_\mathrm{min},\theta_\mathrm{max}]=[\frac{\pi}{2}-h_p,\frac{\pi}{2}]$ (hence
we simulate only one hemisphere of the disc and use reflecting
boundary conditions at the midplane).  Finally our azimuthal extent is
$[\phi_\mathrm{min},\phi_\mathrm{max}]=[-\frac{\pi}{18},\frac{\pi}{18}]$. The cell
numbers in each direction are
$(N_r,N_{\theta},N_{\phi})=(880,64,1536)$. Recast in pressure
lengthscale our box size is $(4H, H, 7H)$, while its resolution is
$(2.3\cdot 10^{-4}r_p, 7.8\cdot 10^{-4}r_p,2.3\cdot 10^{-4}r_p)$. To
avoid reflections on the radial boundaries of our computational box,
we use damping boundary conditions as in \citet{valborro06}, the width
of the inner damping ring being $3\cdot 10^{-2}r_p$ and that of the
outer ring being $3.5\cdot 10^{-2}r_p$,  the damping timescale at
the edge of each damping ring being $1/20^{th}$ of the local orbital period.

The size of the box considered is too small to properly capture the
Lindblad and corotation torques. However, when our focus is on the
heating torque, we subtract a ``cold run'' from a ``hot run'', i.e. a
run without energy release by the planet from a run with a luminous
planet. Our set of parameters is such that the flow is weakly perturbed
over the whole mesh, so that the perturbation arising from the
planet's gravity is nearly independent of that due to its
luminosity. By subtracting both runs we are therefore essentially left
with the thermal disturbance, which fits comfortably in our
computational domain.

Similarly, when our focus is on the cold thermal torque, we subtract
an adiabatic run from a ``cold run'', i.e. a run with a non-luminous
planet and thermal diffusion. Again, we are left essentially with
a thermal disturbance of size $\lambda_c$, which is properly captured
on our domain. Arguably thermal diffusion may also alter the Lindblad
torque, and the corotation torque under some circumstances. We will
discuss these issues later.

In all our simulations, the planet is held on a fixed circular orbit
and is located at the intersection between cell interfaces in azimuth,
radius and colatitude, so that it lies at the centre of an
eight-cell cube. One eighth of the energy released is added to these
cells at each timestep, as in \citet{2015Natur.520...63B}.

\section{Summary of results from linear theory}
\label{sec:summary-results-from}

The expressions for the cold and heating thermal torques are
respectively \citep{2017MNRAS.472.4204M}:
\begin{equation}
\Gamma_{\mathrm{thermal}}^{\mathrm{cold}}=-1.61\frac{\gamma-1}{\gamma}\frac{x_p}{\lambda_c}\Gamma_0
 \label{eq:cold}
\end{equation}
and 
\begin{equation}
\Gamma_{\mathrm{thermal}}^{\mathrm{heating}}=1.61\frac{\gamma-1}{\gamma}\frac{x_p}{\lambda_c}\frac{L}{L_c}\Gamma_0,
 \label{eq:heating}
\end{equation}
where
\begin{equation}
L_c=\frac{4\pi GM_p\chi\rho_0}{\gamma},
 \label{eq:luminosity}
\end{equation}
$\rho_0=\frac{\Sigma_0}{\sqrt{2\pi}H}$ being the unperturbed
midplane density at $r=r_p$, and where
\begin{equation}
\Gamma_0=\Sigma r_p^4\Omega_p^2\left(\frac{M_p}{M_\star}\right)^2h^{-3},  
 \label{eq:gamma0}
\end{equation} 
Eqs.~\eqref{eq:cold} and~\eqref{eq:heating} have been
obtained in the limit $|x_p| \ll \lambda_c$. The distance $x_p$ of the
planet to its corotation is given by:
\begin{equation}
x_p=\eta h_p^2r_p,
 \label{eq:corotation}
\end{equation}
where  $\eta$ is a function of the density and temperature gradients:
\begin{equation}
\eta=\frac{\alpha}{3}+\frac{\beta+3}{6}.
\end{equation}
The size $\lambda_c$ of the thermal disturbance is given by:
\begin{equation}
\lambda_c=\sqrt{\frac{\chi}{(3/2)\Omega_p\gamma}}.
 \label{eq:lambda}
\end{equation}
The sum of equations (\ref{eq:cold}) and (\ref{eq:heating}) gives the
net or total thermal torque:
\begin{equation}
\Gamma_{\mathrm{thermal}}^{\mathrm{total}}=1.61\frac{\gamma-1}{\gamma}\frac{x_p}{\lambda_c}\left(\frac{L}{L_c}-1\right)\Gamma_0.
 \label{eq:total_thermal}
\end{equation}
As said above, these expressions are valid in the limit
$x_p/\lambda_c\rightarrow 0$. When, on the contrary the distance to
corotation is large compared to the size of the disturbance, the force
acting on the planet should tend to that worked out by
\citet{2017MNRAS.465.3175M}, which has the expression:
\begin{equation}
  \label{eq:5}
  F^\infty_\mathrm{heat}=\frac{\gamma(\gamma-1)GM_pL}{2\chi c_s^2},
\end{equation}
where $c_s$ is the adiabatic sound speed and where the $\infty$ sign
conveys the fact that the distance to corotation is large in units of
$\lambda_c$. Eq~\eqref{eq:5} gives the heating force acting on a
perturber with low Mach number in a medium with negligible shear. Note
that this force can only be attained if the Keplerian flow can still
be largely subsonic for $x_p\gg \lambda_c$, i.e. if there exists
$x_p\gg \lambda_c$
such that we have $\Omega_px_p\ll c_s$. The force of Eq.~\eqref{eq:5}
is therefore attained if the corotation offset satisfies the double
hierarchy:
\begin{equation}
  \label{eq:8}
  \lambda_c\ll x_p\ll H.
\end{equation}
Using
Eqs.~\eqref{eq:luminosity}, \eqref{eq:gamma0} and the relationship
between the adiabatic sound speed $c_s$ and the pressure lengthscale
which reads:
\begin{equation}
  \label{eq:6}
  H=\frac{c_s}{\gamma\Omega},
\end{equation}
we can write the corresponding, asymptotic torque $r_pF^\infty_\mathrm{heat}$ as:
\begin{equation}
  \label{eq:7}
  \Gamma^\infty=\sqrt{2\pi}\frac{\gamma-1}{\gamma}\frac{L}{L_c}\Gamma_0.
\end{equation}
Therefore, for a large planet offset, the heating torque normalized to
$\Gamma_0(L/L_c)$ should tend toward the constant value
$\sqrt{2\pi}\frac{\gamma-1}{\gamma}\approx 0.72$ (for $\gamma=7/5$ as here) or
$1.00$ (for $\gamma = 5/3$). 

\section{Heating torque}
\label{sec:heating-torque}

In this section we study numerically the density perturbation arising 
from the release of heat by the planet and its diffusion in the disc, for the case of a planet centred on corotation then for the case of a planet offset from corotation. The runs used
to obtain the maps of perturbation of surface density are performed using the softening length $\epsilon_1$. Next we turn to
the impact of the perturbation of density on the torque. As explained in 
section~\ref{sec:code-mesh-domain}, for each set of parameters we 
perform two runs: one with a luminous planet, one with the planet's 
luminosity switched off, and subtract the results.


\begin{figure*}
  \centering
   \includegraphics[width=.49\textwidth]{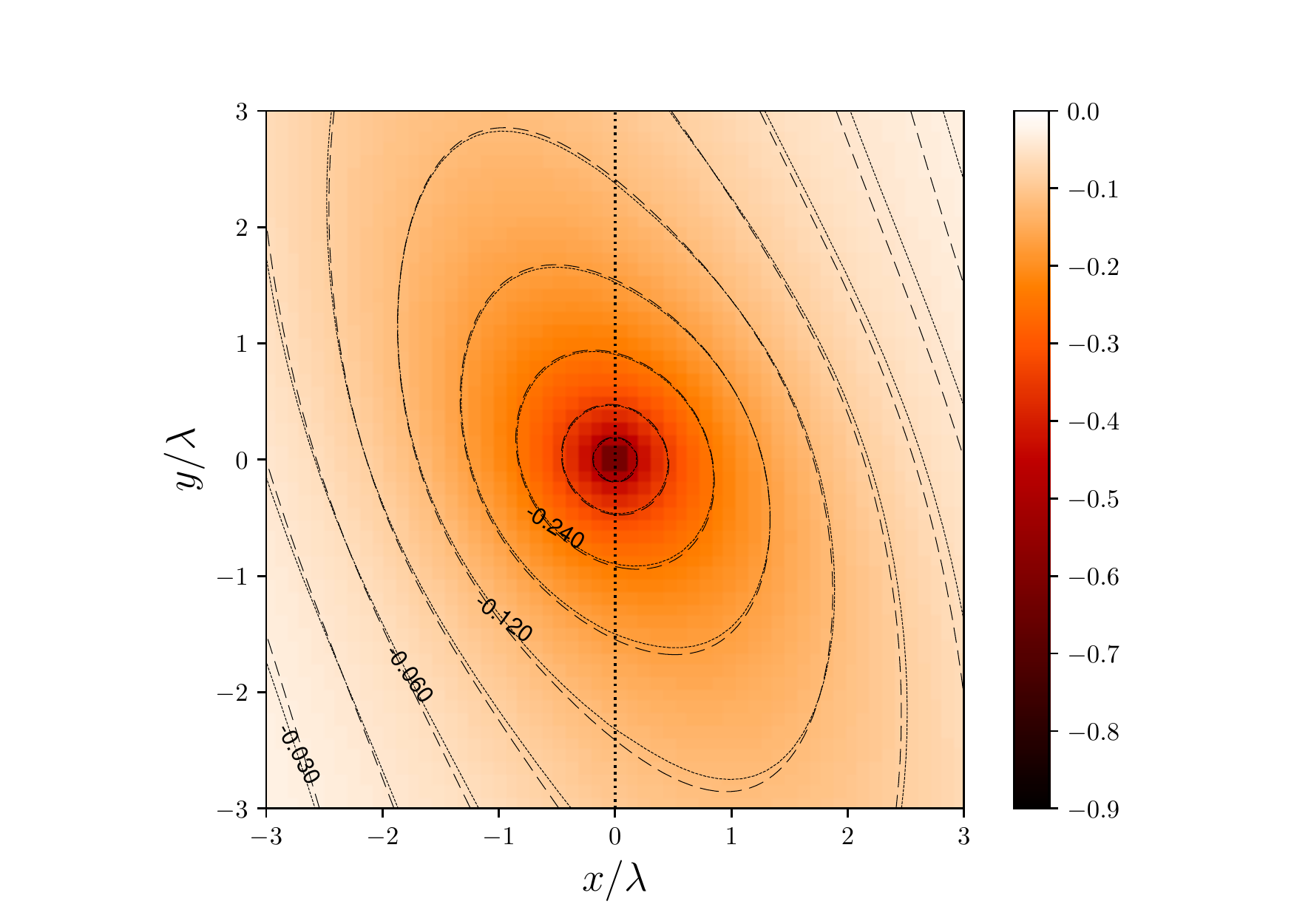}~%
   \includegraphics[width=.49\textwidth]{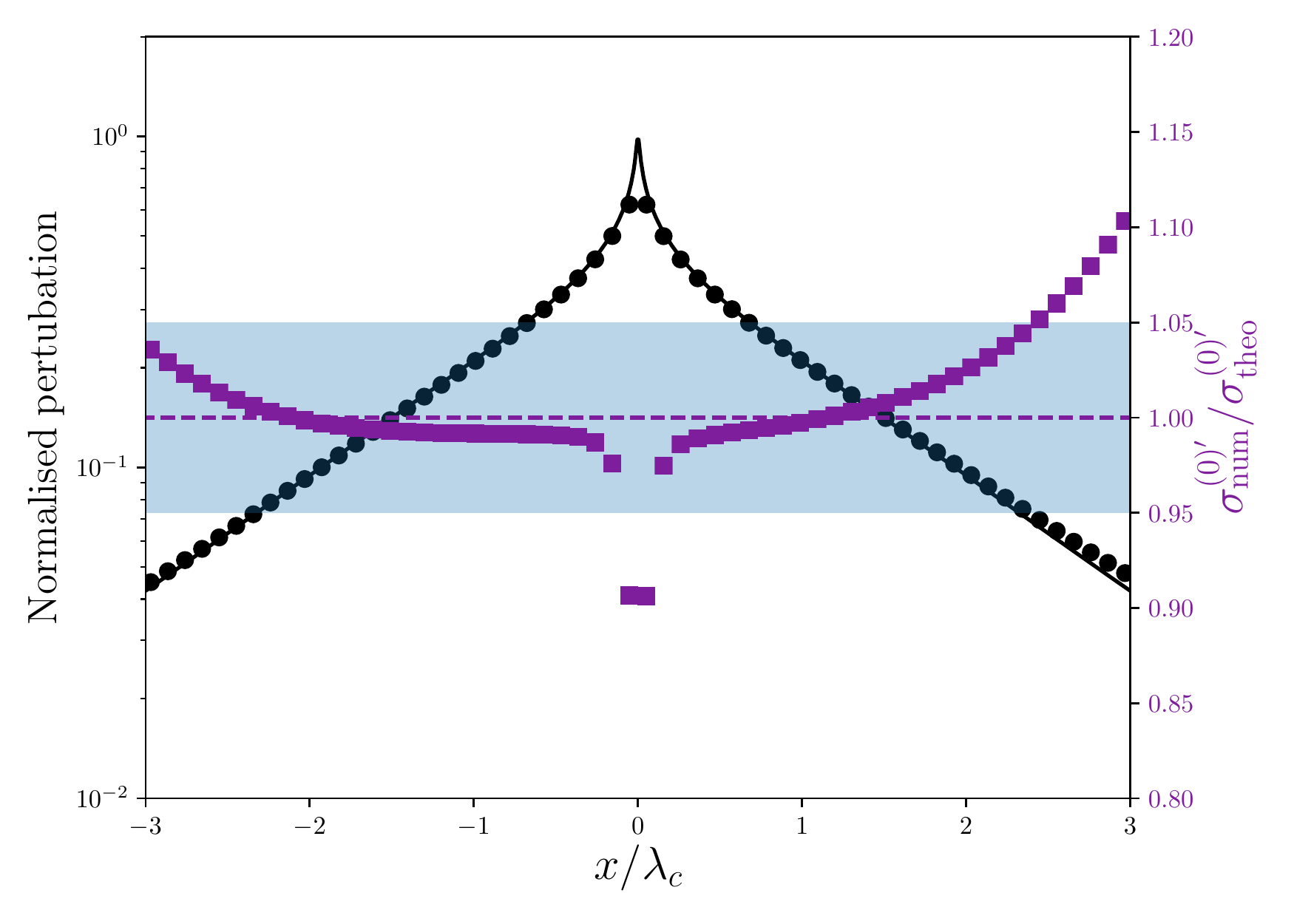}
   \caption{\label{f:density}Left: perturbation of surface density
     $\sigma'^{(0)}$ normalized to $\gamma(\gamma-1)L/\chi
     c^2_s$. Solid lines:  isocontours of the fields obtained after 
    $14$~orbital periods of the planet. Dashed lines: isocontours 
    of the perturbation predicted by linear theory. Here $x=r-r_p$ and 
    $y=r_p\phi$. This map is obtained by subtracting two runs (a hot
    one and a cold one) with a planet centred on corotation, as
    described in section~\ref{sec:case-planet-centred}. Right: cut of
    the absolute value of 
    $\sigma'^{(0)}$ along the axis $y=0$.  The black circles represent
    the values measured in the simulation, while the solid black line
    represents the value expected from linear theory.  The squares (in
  purple in the electronic version of this manuscript) represent the
  ratio of the value measured in the numerical simulation to the value
expected from linear theory. The light horizontal band delineates the
region where these values coincide to within $\pm 5$~\%.}
\end{figure*}

\begin{figure*}
  \centering
   \includegraphics[width=.49\textwidth]{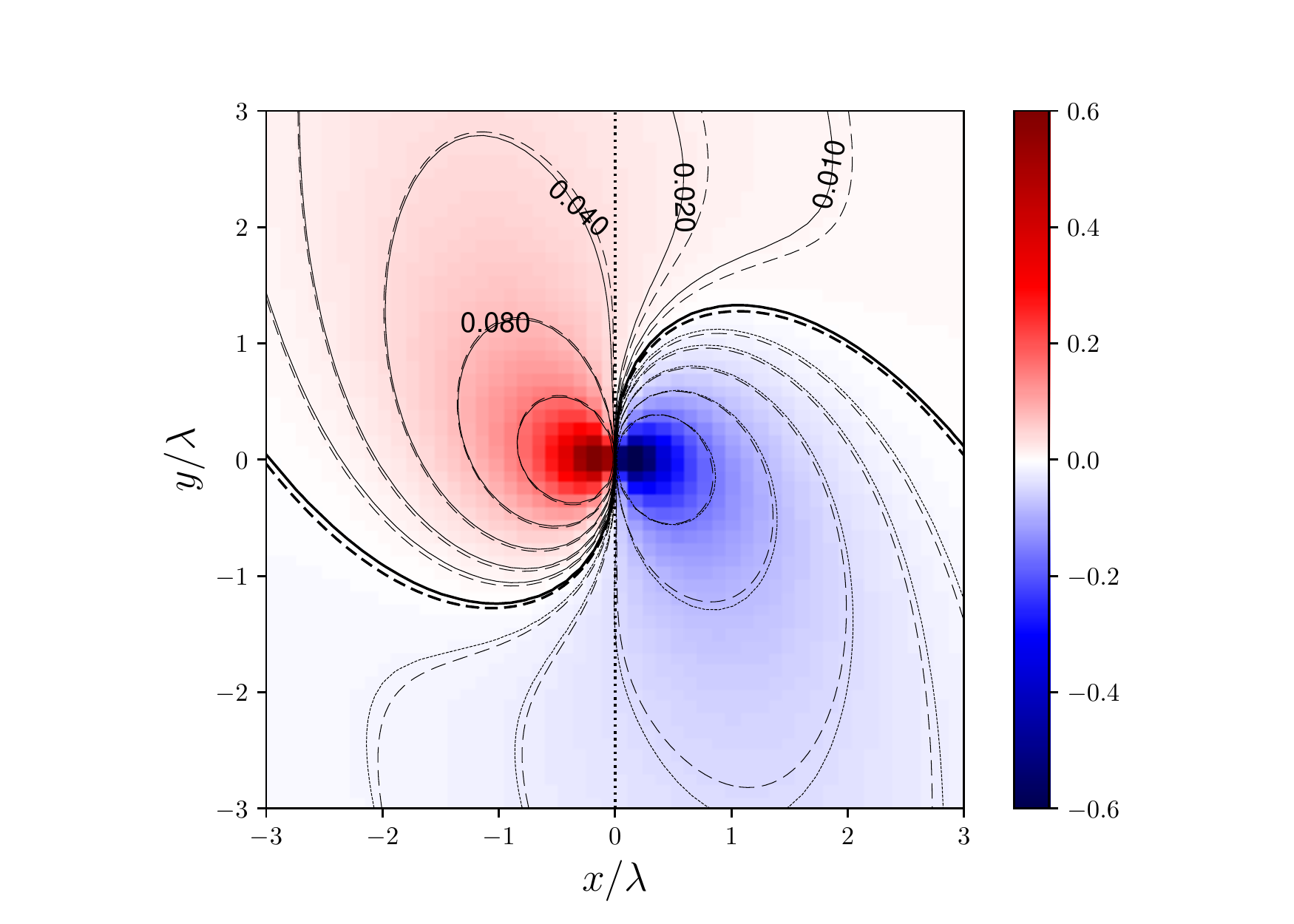}~%
   \includegraphics[width=.49\textwidth]{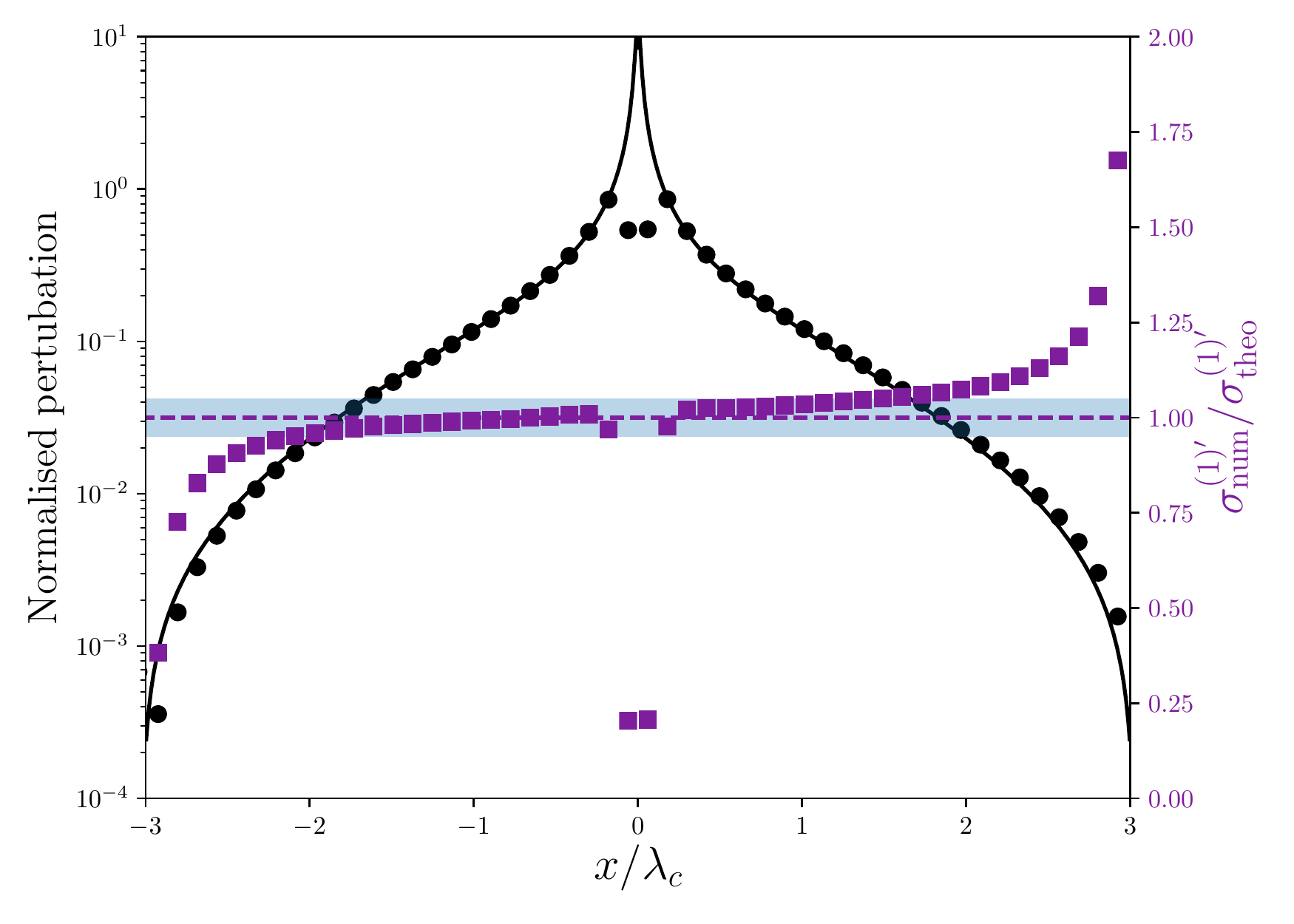}
  \caption{\label{f:derivate}Left: approximation of the partial derivative of the perturbed surface 
  density with respect to $x_p$, $\sigma^{'(1)}$, normalised to 
  $\gamma(\gamma-1)L/(\chi c_s^2\lambda_c)$. The line style is the same
  as that of Fig.~\ref{f:density}. This map has been obtained using
  the four runs described in
  section~\ref{sec:case-planet-offset}. Right: cut of the absolute
  value of the map along
  the $y=0$ axis. The symbols have same meaning as in
  Fig.~\ref{f:density}. The light horizontal band shows again the
  region where numerical and theoretical values coincide to within $\pm 5$~\%.}
\end{figure*}


\begin{figure}
 \centering\includegraphics[width=0.5\textwidth]{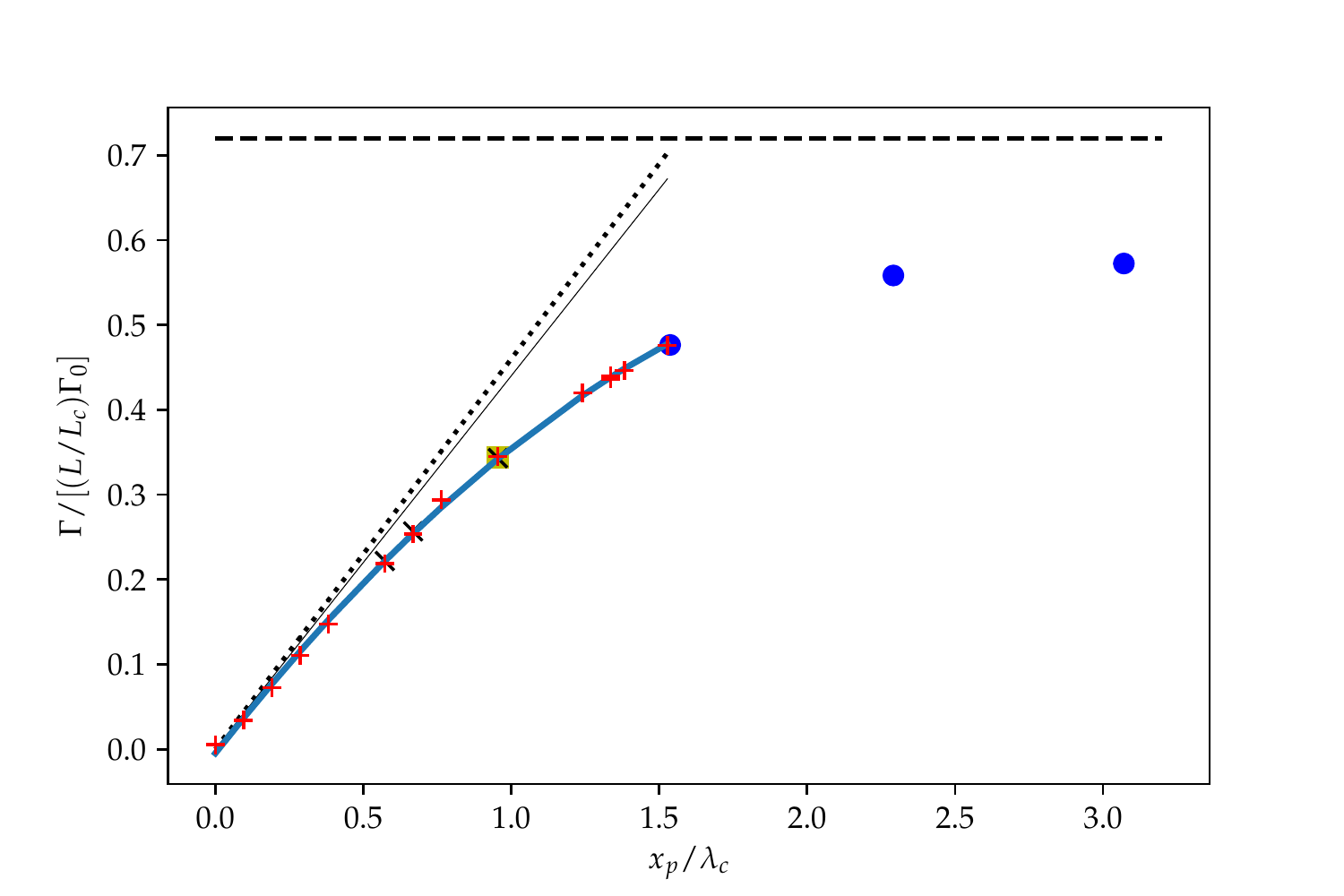}
 \caption{Normalized heating torque as a function of the corotation
   offset. The `+' and `x' symbols, as well as the square symbol,
   represent runs performed with diverse values of $\alpha$ and
   $\beta$. The disc symbols represent additional runs performed with
   a planet outside of its nominal position in a strictly Keplerian
   disc (see text for detail). The `x' and square symbol have been
   used whenever another run had been performed with same value of
   $x_p$, in order to distinguish the symbols. The short dashed line
   shows the expected trend for $x_p/\lambda_c \rightarrow 0$ whereas
   the thin solid line shows the tangent to the polynomial fit in
   $x_p=0$. The horizontal dashed line shows the asymptotic value of
   Eq.~\eqref{eq:7}. The thick blue line (in the electronic version of
   this article) shows a second order polynomial fit to the data
   obtained from sub-Keplerian discs.}
   \label{f:xp}
\end{figure}

\subsection{Case of a planet centred on corotation}
\label{sec:case-planet-centred}
We first consider the case of a planet embedded in a strictly
Keplerian disc that has $\alpha=-2$ and $\beta=1$. In that case, the
planet's orbit and corotation coincide: $x_p=0$, and no torque is
expected from the thermal disturbance. In that case we verify that the
perturbation of surface density corresponds to that predicted by
linear theory for a planet centred on corotation
\citep{2017MNRAS.472.4204M}. We represent on the left plot of
Fig.~\ref{f:density} the integral $\sigma^{'(0)}$ along the
$\theta$-direction of the difference of density fields between a hot
and cold run. This field matches satisfactorily the predictions of
linear theory, both in shape and amplitude: the right plot of
Fig.~\ref{f:density} shows that both fields coincide to within
$\pm 5$~\% over the domain represented, except in the rightmost part
of the map, where the discrepancy reaches $\sim 10$~\%, and at the
innermost cells. At the distance at which the departure between the
two fields exceeds $5$~\%, the field value is about one order of
magnitude smaller than its value at the distance $\lambda_c$ from the
planet. We finally comment that the isocontours from linear
theory, which has been formulated in the shearing sheet formalism, are
symmetric by construction. In contrast, a small asymmetry of the
contours from the numerical simulation is apparent.

\subsection{Case of a planet offset from corotation}
\label{sec:case-planet-offset}
We then turn to the case of a planet offset from corotation. Namely,
we check that the derivative of the perturbation of surface density
with respect to $x_p$, in $x_p=0$, is in agreement with that predicted
by linear theory. For this purpose we study two cases (corresponding
to four runs in total) with two small values of $x_p$, with same
absolute value and different signs. In this section only we used a
resolution in radius of $768$ cells instead of $880$. We take in one
case $(\alpha,\beta)=(-1.6875008,1.0)$, which yields a corotation
offset $x_p^0$  exactly equal to the radial resolution, and in
the other case $(\alpha,\beta)=(-2.3124992,1.0)$, which yields a
corotation offset $-x_p^0$. Prior to subtracting the runs with
different offsets, we shift the maps one cell in the radial direction,
outward for that with an offset of $x_p^0$ and inward for the other
one, so that their corotations are superimposed.

We show in the left plot of
Fig.~\ref{f:derivate} the map
$[\sigma^{(0)}(x_p^0)-\sigma^{(0)}(-x_p^0)]/(2x_p^0)$ and compare it
to analytical expectations.  The agreement between numerics and
analytics is satisfactory, although the mismatch between the actual
and expected isocontours is larger than in the map of
Fig.~\ref{f:density}, as can be seen in the right plot of
Fig.~\ref{f:derivate}. We see that the value from numerical
simulations coincides to within $5$~\% with that expected from linear
theory up to a distance $\sim 1.5\lambda_c$ from the planet (except
for the two innermost cells). At the edge of the region represented,
the relative discrepancy becomes large, while the field is about two
orders of magnitude smaller than at the distance $\lambda_c$ from the
planet.

\subsection{Torque dependency on the gradients of surface density and temperature}
\label{sec:depend-heat-torq}
Linear theory predicts that the heating torque, for a given disc and
planet, only depends on the distance of the planet from
corotation. This distance, in turn, depends in a non unique manner on
the slopes of surface density and temperature $\alpha$ and $\beta$.
Here, by varying these two parameters, we consider a wide range of
values for $x_p$, from zero or largely smaller than $\lambda_c$, to
larger than $\lambda_c$. Note that despite the marked scale separation
between $\lambda_c$ and $H$ (see Eq.~\ref{eq:4}), there is hardly
enough room between these two values to accommodate a value of $x_p$
that simultaneously fulfils the two requirements of scale separation
given by Eq.~\eqref{eq:8}, so we expect the torque estimate at large
corotation offset given by Eq.~\eqref{eq:7} to be only approximate.

On the other hand, in the limit $|x_p |\ll \lambda_c$,
Eq.~\eqref{eq:heating} shows that the heating torque normalized to
$\Gamma_0(L/L_c)$ should be a linear function of $x_p/\lambda_c$ with
slope $1.61\frac{\gamma-1}{\gamma}\approx 0.46$ (for $\gamma=7/5$) or
$0.64$ (for $\gamma=5/3$).

We have performed a variety of runs to check these expectations. In a
first exploration, we have run $18$ cases with values of
$\alpha$ and $\beta$ chosen arbitrarily in the interval $[-2,2]$ interval and
a planet with a luminosity close to the critical luminosity $L_c$
(namely $L=0.965L_c$) and checked that the heating torque is indeed a
one-to-one function of $\eta$.

Fig.~\ref{f:xp} shows the comparison between the results of our
runs and Eqs.~\eqref{eq:heating} and~\eqref{eq:7}. All points fall on
a curve to a high level of approximation, which confirms that the
heating torque does not have a dependency on the gradients of surface
density and temperature other than that borne by the corotation offset.

We also have explored the regime of large corotation offsets with runs
additional to those mentioned above. Achieving a large corotation
offset with a power law disc requires large gradients of surface
density or temperature, which may lead to problems of numerical
stability. Instead, we have explored this regime using a strictly
Keplerian disc (with $\alpha=-2$ and $\beta=1$) and have artificially
moved the planet to larger orbital radii, while keeping its orbital
frequency. In this specific case only, the planet is no longer at the
intersection of cell interfaces, and the energy release is performed
according to the prescription described in
\citet{2017arXiv170401931E}. These extra runs are represented with
filled circles in Fig.~\ref{f:xp}. We have checked the validity of our
method by choosing for the leftmost of these points an offset equal to
the largest one obtained in the runs with varying values of $\alpha$
and $\beta$, and found that the torque thus obtained does indeed
coincide with that obtained from sub-Keplerian discs.

We perform a second order polynomial fit of the data from sub-Keplerian
discs of the form:
\begin{equation}
  \label{eq:9}
  \frac{\Gamma}{(L/L_c)\Gamma_0}=a\left(\frac{x_p}{\lambda_c}\right)^2+b\left(\frac{x_p}{\lambda_c}\right)+c,
\end{equation}
and find a slope at the origin $b\approx 0.44$, about $5$~\% below the
value of $0.46$ expected from analytics. We also find $a\approx
-0.083$ and $c\approx -4.8\cdot 10^{-3}$, this last value being close
to zero, as expected for a vanishing corotation offset.

There is a slight dispersion of the torque measurements with respect
to the polynomial fit. This may be due to a dependence of the heating
on the disc gradients other than that borne by the corotation
offset. These effects are minute, however: we find that the residual
has an r.m.s of $4.5\cdot 10^{-3}\Gamma_0(L/L_c)$, with extreme values
of $(^{-5}_{+10})\cdot 10^{-3}\Gamma_0(L/L_c)$.

\subsection{Convergence study}
\label{sec:conv-resol-study}
We analyse the effect of the mesh resolution on the heating torque by
varying the number of zones in all directions from $0.1$ to $0.9$
times that of the fiducial resolution, by steps of $0.1$, while
maintaining the size of the box. In addition, we consider an extra
resolution of $1.2$ times the fiducial one. As our original number of
zones in radius is a multiple of $20$, the radial number of zones is
an even number in all our cases. This is not necessarily true for the
number of zones in azimuth. When it is odd, the planet is at the
centre of a zone in azimuth, and the heat is then released in a
four-zone layer instead of within an eight-zone cube, following the
general prescription of \citet{2017arXiv170401931E}. 

We consider two values for the ratio $x_p/\lambda_c$: $0.15$ and
$0.95$, that are realised with the pairs of values
$(\alpha,\beta)=(-1.607,1.0)$ and $(\alpha,\beta)= (0.5,1.0)$,
respectively.  The lowest of these two ratios satisfies reasonably
well the requirement $x_p\ll \lambda_c$. However, the value of the
torque in this case is small, and may slightly differ from that given
by linear theory owing to the slight dispersion mentioned in the
previous section.

Conversely, the largest of these two ratios falls in the regime where
the torque squarely departs from the analytic estimate obtained
assuming $x_p\ll \lambda_c$. In this case, the residual dispersion
with respect to the polynomial fit is a smaller fraction of the
torque, but there is no analytic formula that predicts what the torque
value should be.

\begin{figure}
 \centering\includegraphics[width=0.5\textwidth]{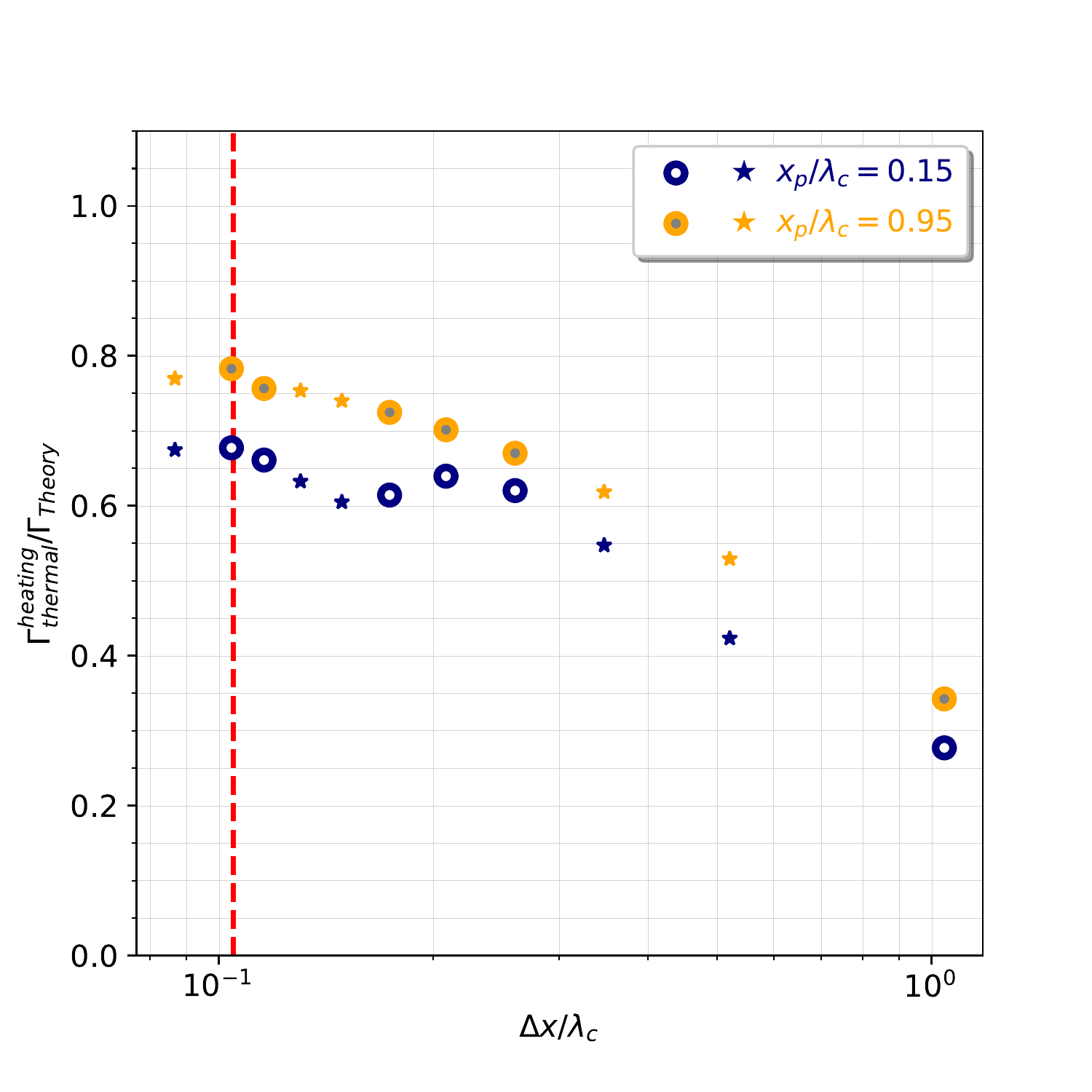}
 \caption{Normalized torque value as a function of the resolution
   normalized to $\lambda_c$, for two different values of
   $x_p$. Circles represent the cases for which the azimuthal number
   of zones is even whereas the star-like symbols represent the cases for
   which this number is odd (which entails a different position of the
   planet with respect to its neighbouring zones and a different
   prescription for energy release, see text for details). The
   vertical, red dashed line represents our fiducial resolution.}
   \label{f:tq}
\end{figure}

Figure (\ref{f:tq}) shows the value of the heating torque normalized
to the theoretical value (Eq.~\ref{eq:heating}) as a function of the
radial resolution $\Delta x$ normalized to the size of the disturbance
$\lambda_c$.  The vertical, red dashed line represents our fiducial
resolution, which resolves $\lambda_c$ over 9.6 cells in radius and
azimuth and 2.8 cells in colatitude.
We can see that the maximum value of the heating torque obtained from
the simulations for the case when $x_p/\lambda_c=0.15$ is
$\approx 70$~\% of the value predicted by linear theory, while it is
$\approx 80$~\% of the theoretical value for $x_p/\lambda_c=0.95$.  We
also see that the torque value does not vary monotonically with the
resolution.  In particular, the maximum value for the heating torque
is obtained for the fiducial resolution. When we increase the
resolution in all directions by the factor $1.2$, we obtain a value
slightly smaller than that obtained for the fiducial resolution, for
the two values of $x_p/\lambda_c$ considered.

The fact that the torque value is $\approx 20$~\% smaller than the
theoretical prediction for $x_p/\lambda_c=0.95$ is largely
attributable to the fact that $x_p$ is not small compared to
$\lambda_c$. Using the second order fit performed in
section~\ref{sec:depend-heat-torq}, and noting
$X\equiv x_p/\lambda_c=0.95$, we have $(aX^2+bX)/bX\approx 0.82$ (note
that we discard the small value of the constant term $c$ here): the
expected value is approximately $20$~\% smaller than the linearly
extrapolated value $bX$. This fit has been obtained for the finite,
fiducial resolution, but it is reasonable to expect that the drop
between the linearly extrapolated value and the actual value is
relatively independent of resolution and of the order of $20$~\%.

The reason for the discrepancy between numerics and the theoretical
formula in the other case ($x_p/\lambda_c=0.15$) is less clear. Here,
the expected value of the torque is only
$0.46\times 0.15\Gamma_0\approx 0.07\Gamma_0$: this small value
compounds any minute variation that the torque may have with respect
to the analytic expectation. A mismatch of $30$~\% on the torque value
therefore suggests that effects neglected in the formulation of
\citet{2017MNRAS.472.4204M}, such as the dependence on the temperature
gradient or curvature effects might be as large as $0.02\Gamma_0$,
comparable in magnitude to, but larger than the maximal dispersion
with respect to the polynomial fit that we measured in
section~\ref{sec:depend-heat-torq}.

Anticipating on the behaviour at larger mass that will be explored in
section~\ref{sec:toward-larger-masses}, we also note that the mass
considered in this study is only a factor of $\sim 6$ smaller than the
critical mass beyond which thermal effects are cut off
\citet{2020arXiv200413422V}. Although it is unlikely that such a small
mass would result in a $30$~\% cut off, it may still account for a few
percents difference with respect to the value expected from linear
theory.

We also note
that for this case, the corotation offset $x_p$ is just marginally
larger than the mesh resolution and the softening length even at the
largest resolution (the softening length $\epsilon_2$ scales with the
resolution in this convergence study). A study with resolution even
higher than that considered in the present work is warranted to
disentangle finite resolution effects from additional, minor terms in
the expression of the heating torque.

\section{Full torque}
\label{sec:cold-thermal-torque}
\begin{figure}
 \centering\includegraphics[width=\columnwidth]{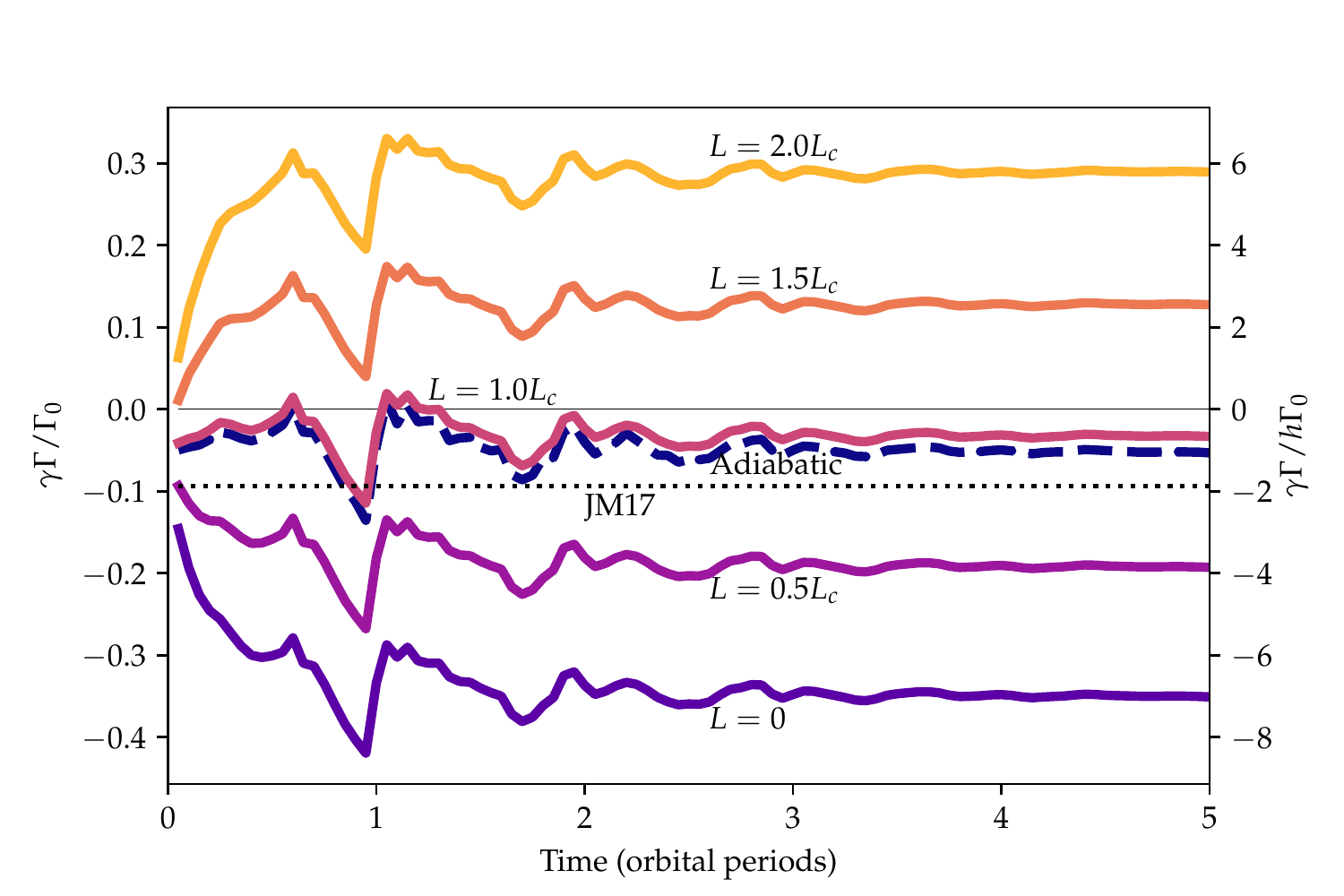}
 \caption{Normalized total torque as a function of time for different
   cases. The dashed curve shows the torque  in the adiabatic
   disc described in the text. The thick lines show the torque in a  disc with
   similar parameters, except that is has a finite thermal
   diffusivity, for different planetary luminosities: the bottom curve shows the torque for a non-luminous
   planet, while the other curves show the torque for a planet of increasing
   luminosity, from $0.5L_c$ to $2L_c$ by steps of $0.5L_c$, from
   bottom to top.
   The horizontal dotted line shows the torque value expected in the
   adiabatic disc according to \citet{2017MNRAS.471.4917J}. The right
   vertical axis corresponds to the usual normalization of the torque.
   \label{f:ct0}}
\end{figure}
We now turn to the case of the  total torque. We consider a disc
with same parameters as those described in
section~\ref{sec:initial-conditions}, with $\alpha =0$ and $\beta=0$,
and a planet of mass $10^{-7}M_\star$ as previously. We have
therefore in such disc:
\begin{equation}
  \label{eq:10}
  \frac{x_p}{\lambda_c}\approx 0.57
\end{equation}
We perform
several runs in which we vary the planet's luminosity from $0$ to
$2L_c$. We also perform an additional run with a non-luminous planet
in an adiabatic disc (with exact same parameters except that it has
$\chi=0$). We plot in Fig.~\ref{f:ct0} the torque obtained in the
different runs as a function of time, over the first $5$~orbital
periods.

The torque are presented with two different normalization: in terms
of $\Gamma_0$ on the left axis, which naturally arises from the
formulation of Eqs.~\eqref{eq:cold}-\eqref{eq:heating}, and in terms
of $\Gamma_0h$ on the right axis\footnote{Following
  \citet{2017MNRAS.472.4204M} we use for $\Gamma_0$ the scaling
  $\Sigma\Omega_p^2r_p^4q^2h^{-3}$ of the one-sided Lindblad torque
  instead of the more commonly used reference
  $\Gamma_0h=\Sigma\Omega_p^2r_p^4q^2h^{-2}$ which scales with the
  differential Lindblad torque.}, which follows the scaling of
Lindblad and corotation torques. In addition, as is customary in the
literature, we multiply the torque by the adiabatic index
$\gamma$. This figure helps to grasp the importance of thermal
torques. When those are not present, the normalized torque (right
axis) is usually comprised between $\sim -2$ and $\sim +2$
\citep[e.g.][]{2015MNRAS.452.1717L}, whereas here it spans a much
broader range from $\sim -7$ to $\sim +6$.

We show on that figure the torque expected in the adiabatic disc,
according to the torque formulae of \citet{2017MNRAS.471.4917J}. Not
surprisingly, the torque that we measure is smaller: it is not
possible to correctly capture the Lindblad and corotation torques with
our small computational domain. However, thermal disturbances largely
fit within our domain, so that our estimate of thermal torques, obtained by
subtraction of two runs, is not affected by this effect.

We see that the different curves display an offset nearly constant in
time between each other past one orbital period: this is the time it
takes to establish thermal torques. We also see that the torque
obtained for $L=L_c$ is nearly identical to the torque in an adiabatic
disc, as can be expected from Eqs.~\eqref{eq:cold}
and~\eqref{eq:heating}: for $L=L_c$, thermal torques are expected to
cancel out.

\begin{figure}
 \centering\includegraphics[width=\columnwidth]{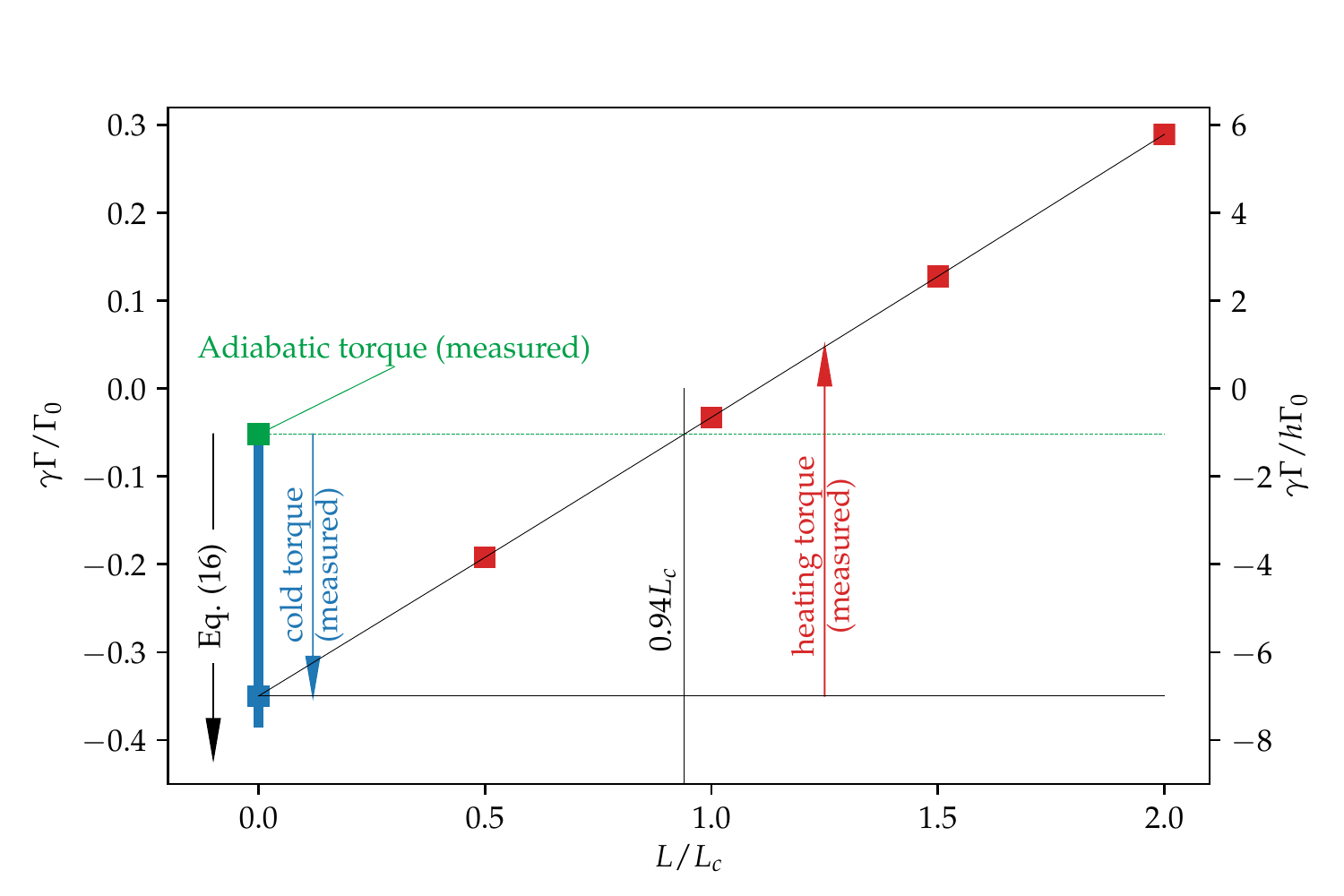}
 \caption{Torque measured in the different runs mentioned in the text
   (square symbols). In the electronic version, the colour matches that
   of \citet{2019MNRAS.483.4383V}: we use green for adiabatic
   calculations, blue for cold runs and red for hot runs. The torque
   normalisation on both vertical axis are the same as those of
   Fig.~\ref{f:ct0}. All measurements of the torques correspond to the time
   averaged value between the 4\textsuperscript{th} and
   5\textsuperscript{th} orbital period. The thick vertical line at
   $L=0$ (in blue in the electronic version) has a length given by
   Eq.~\eqref{eq:cold} multiplied by $1+ax_p/b\lambda_c$
   (see text for detail).
   \label{f:ct1}}
\end{figure}

Fig.~\ref{f:ct1} shows the time averaged value of the torques of
Fig.~\ref{f:ct0} as a function of the planetary luminosity. We see
that the cold thermal torque measured in our runs is in correct
agreement with the value predicted by Eq.~\eqref{eq:cold}. Although we
do not have undertaken a systematic study of this torque as a function
of $x_p/\lambda_c$ as we did in section~\ref{sec:heating-torque} for
the heating torque, it is natural to expect that this torque follows the
same deviation from the proportionality law in $x_p/\lambda_c$, since
the diffusion and advection equation governing the evolution of the
thermal lobes is identical for the cold and heating torques
\citep{2017MNRAS.472.4204M}, and since a non-luminous planet with a
large corotation offset is subjected to the force given by
Eq.~\eqref{eq:5} with $L=-L_c$
\citep{2019MNRAS.483.4383V,2020arXiv200413422V}.  For the not so small
value of $X\equiv x_p/\lambda_c$ used in our calculations (see
Eq.~\ref{eq:10}), we expect a drop of the cold thermal torque
of order $(aX^2+bX)/bX=1+ax_p/b\lambda_c\approx 0.89$. We
report the value of the expected cold thermal torque thus corrected in
Fig.~\ref{f:ct1} as a thick vertical line. Our measurement is
marginally smaller, by $\sim 10$~\%, than this expected value.

We also report in Fig.~\ref{f:ct1} the value of the luminosity
required for the thermal torques to cancel out (i.e. for the net
torque to be equal to the adiabatic torque). We find $L=0.94L_c$. The
fact that this value is not exactly $L_c$ as expected from
Eqs.~\eqref{eq:cold} and~\eqref{eq:heating} can be attributable to the
resolution used in our runs, and also possibly to the fact that the
introduction of a finite thermal diffusivity in the disc yields
corrections to the Lindblad torque in addition to the appearance of
the cold thermal torque \citep{2017MNRAS.472.4204M}. These corrections
should be minute, however, as the introduction of a finite thermal
diffusivity induces a torque change in good agreement with that
expected from the cold thermal torque.

\section{Toward larger planetary masses}
\label{sec:toward-larger-masses}
We have seen in the previous sections how numerical estimates of
thermal torques are in reasonable agreement with analytical
expectations. For this purpose, we have used a very small planetary
mass (about one third of the mass of Mars, for a solar-like central
star), which allowed a clean comparison with linear theory. In order
to incorporate thermal torques in models of planetary formation and
migration, one needs reliable torque expressions in the mass range for
which it is crucial to correctly predict migration, i.e. in the Earth
mass range and above.  There is a critical mass that features
prominently in analysis of thermal forces, which reads:
\begin{equation}
  \label{eq:11}
  M_c=\frac{\chi c_s}{G}.
\end{equation}
When the perturber's mass is larger than this critical mass, not all
the energy released by the perturber contributes to heat its
surroundings, resulting in thermal forces smaller than their nominal
values \citep{2020arXiv200413422V}.  Differently said, if the thermal
diffusivity is much smaller than $GM_c/c_s$, the flow in the planet's
vicinity tends to behave adiabatically and thermal forces become
unimportant. \citet{2020arXiv200413422V} have investigated in detail
the cut-off of thermal forces (both the cold force and the heating
force) in the context of (negative) dynamical friction. No such study
has been undertaken in the context of the present work, that of
planets on circular orbits in a sheared flow. Given the considerable
numerical endeavour that such study represents, as one needs to
resolve adequately the Bondi sphere of low mass objects, we do not
present here a systematic study of thermal torques for objects in the
Earth mass range. Instead, we discuss some of the potential
complications that arise in this mass range, and illustrate them with
selected runs.

\subsection{Interaction with the horseshoe
  flow: scaling laws}
\label{sec:crit-mass-plan}
The perturbation of density in the lobes that leads to the torque
expressions of section~\ref{sec:summary-results-from} arises from an
advection-diffusion equation, where advection occurs with the
unperturbed Keplerian flow. When the perturbation of velocity in the
vicinity of the planet is comparable to that of the unperturbed flow
over the length scale of the thermal disturbance $\sim\Omega_p\lambda_c$, one
can expect a significant distortion of the thermal lobes, and
consequently a change in the torque.
The speed at which horseshoe U-turns of low mass 
planets  are executed is, in order of magnitude \citep{bm08}: 
\begin{equation}
  \label{eq:14}
  v_U\sim \frac{GM_p}{\Omega_pH^2}\sim R_b\Omega_p. 
\end{equation}
A significant change in the shape of the thermal lobes and the torque
they exert on the planet is therefore to be expected when:
\begin{equation}
  \label{eq:15}
  v_U\gtrsim \Omega_p\lambda_c 
\end{equation}
which translates into, using Eqs.~\eqref{eq:11} and~\eqref{eq:14}: 
\begin{equation}
  \label{eq:16}
  M_p\gtrsim\frac{c_s^2}{G}\sqrt\frac{\chi}{\Omega}\sim M_c\frac{H}{\lambda_c} 
\end{equation}
We note that this critical mass is not the planet mass for which the
width of the horseshoe region, which is $r_p\sqrt{M_p/(M_\star h)}$
for a low-mass planet \citep{2015MNRAS.452.1717L,2016ApJ...817...19M},
becomes comparable to the extent of the thermal disturbance. This
occurs when:
\begin{equation}
  \label{eq:13}
  M_p\sim \frac{\chi M_\star c_s}{\Omega_p^2r_p^3}\sim M_c.
\end{equation}
In other words, when a planet reaches the critical mass of
Eq.~\eqref{eq:11}, the radial size of the horseshoe region becomes
comparable to that of the thermal disturbance. Yet Eq.~\eqref{eq:16}
shows that planetary masses at least a factor of $H/\lambda_c$ larger
are required to yield a significant distortion of their thermal
lobes. The width of the horseshoe region refers to the width reached
at large azimuthal distance from the planet. At small distances
corresponding to the size of thermal lobes, the horseshoe region is
considerably more narrow. The horseshoe region of a planet that
marginally fulfils $M_p\gtrsim M_c$ is therefore considerably more
narrow, near the planet, than the thermal disturbance, and the lobes
are not significantly distorted.

Noting that the critical mass of Eq.~\eqref{eq:11} can also be written
as $M_\mathrm{th}(\lambda_c/H)^2$,
where $M_\mathrm{th}$, the thermal mass, is given by:
\begin{equation}
  \label{eq:17}
  M_\mathrm{th}=M_\star h^3=\frac{c_s^3}{G\Omega},
\end{equation}
we can enumerate the following set of
cases, in which for brevity we denote with $\theta\equiv \lambda_c/H$ the
(generally small) dimensionless ratio of thermal to pressure length scales:
\begin{enumerate}
\item If $M_p \ll \theta^2M_\mathrm{th}$, thermal torques have a value
  compatible with that given by linear theory
  (Eqs.~\ref{eq:cold}-\ref{eq:heating}).
\item If $\theta^2M_\mathrm{th} < M_p < \theta M_\mathrm{th}$, thermal
  torques are below the value predicted by linear theory, as they are
  in the cut-off regime. The extent of the horseshoe region is
  comparable to or larger than that of the thermal disturbance, but
  the perturbation of velocity of the horseshoe flow should not
  significantly alter the shape of the lobes, so that the magnitude of
  the thermal torque should be that given by linear theory, reduced by
  a cut-off factor of magnitude similar to that investigated by
  \citet{2020arXiv200413422V}.
\item If $\theta M_\mathrm{th} < M_p < M_\mathrm{th}$, the thermal
  torques are significantly cut-off. Besides, the horseshoe flow has a
  magnitude comparable to or larger than that of the unperturbed disc
  over the thermal disturbance, with a strong impact both on the shape
  of the lobes and the magnitude of the thermal torques. Depending on
  the value of $\theta$ and the value of the planet's luminosity,
  thermal torques may be irrelevant compared to the Lindblad and
  corotation torque in this mass regime.
\item Finally, the case of a planetary mass in excess of the thermal
  mass is beyond the scope of this work. It has been investigated
  thoroughly in the literature. In this regime thermal torques
  should be irrelevant.
\end{enumerate}

\subsection{Some illustrative runs}
\label{sec:some-illustr-runs}
\begin{figure}
  \includegraphics[width=\columnwidth]{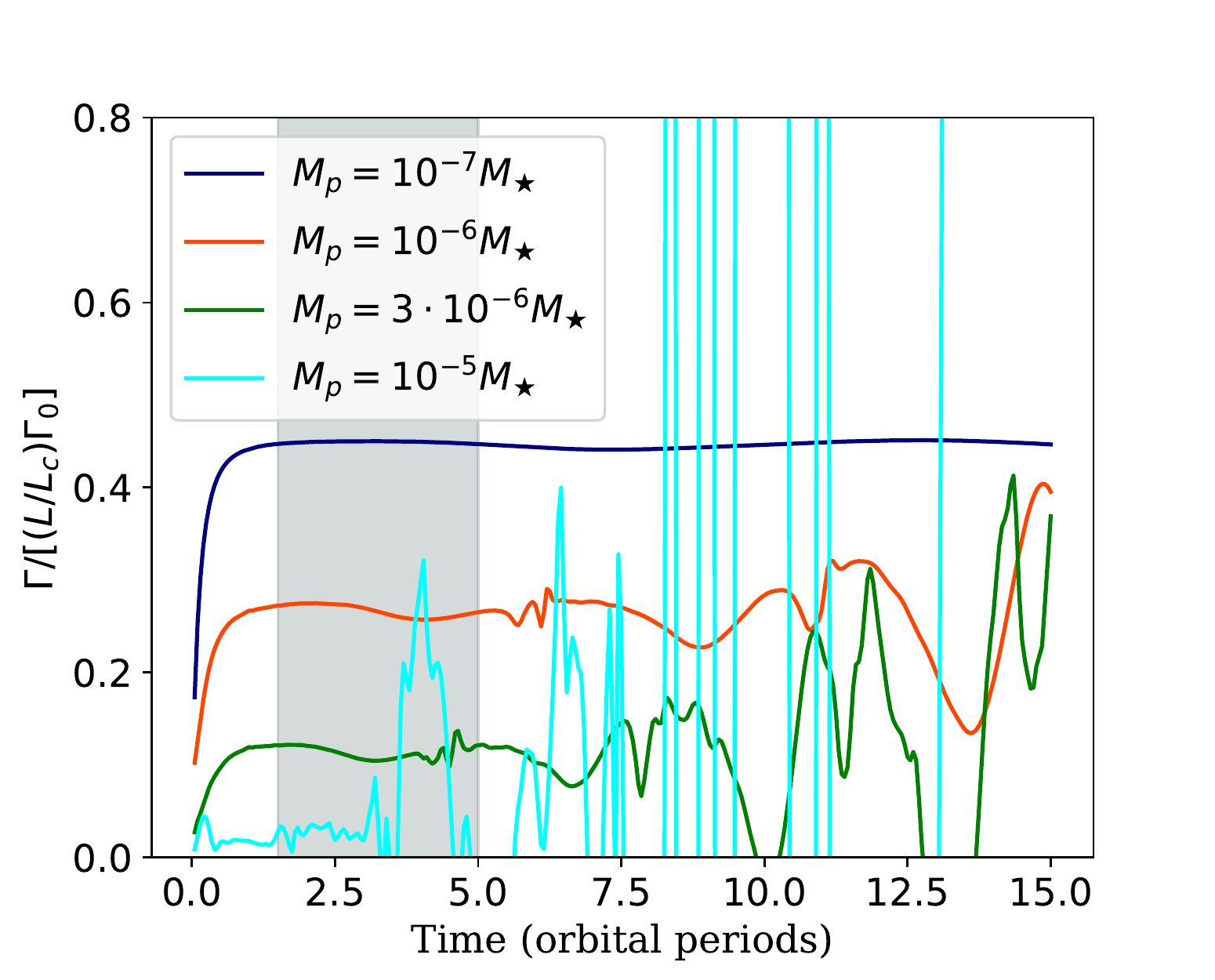}
  \caption{Heating torque normalized to $(L/L_c)\Gamma_0$ as a
    function of time for planets of four different masses.  At larger time the torque fluctuates
    for the planets of larger mass.\label{fig:tqhm}}
\end{figure}

\begin{figure*}
  \includegraphics[width=\textwidth]{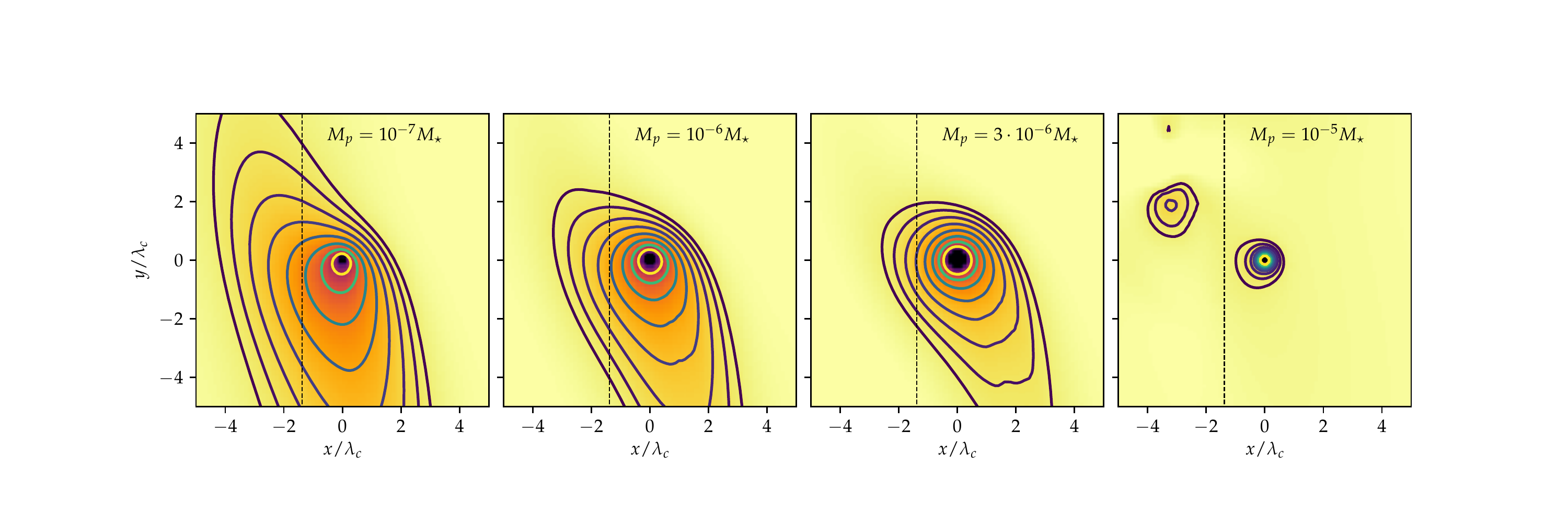}
  \caption{Perturbation of density arising from heat release (obtained
    by subtracting a hot run and a cold run), integrated in colatitude
    and normalized to $\gamma(\gamma-1)L/\chi c_s^2$, at $t=2$~orbits,
    for the four planet masses considered in the text. The vertical
    dashed line shows the corotation. The isocontours have same value
    in the four plots. They start at $0.01$ and are in geometric
    sequence with ratio $\sqrt 2$.\label{fig:lbhm}}
\end{figure*}

We present here four runs which fall in the different regimes
detailed above. The disc parameters are the same than those of
section~\ref{sec:initial-conditions}, so that we have: $\theta\approx
0.043$, $M_\mathrm{th}=1.25\cdot 10^{-4}\;M_\star$, hence we have:
\begin{eqnarray*}
  \theta M_\mathrm{th} &=& 5.4\cdot 10^{-6}\;M_\star\\
  \theta^2 M_\mathrm{th} &=&2.4\cdot 10^{-7}\;M_\star.
\end{eqnarray*}
We consider planet masses of $10^{-7}\;M_\star$ [which falls
marginally in the regime (i) of section~\ref{sec:crit-mass-plan}],
$10^{-6}\;M_\star$ [which is on the lower side of regime (ii)],
$3\cdot 10^{-6}\;M_\star$ [which falls rather on the upper side of
regime (ii)] and $10^{-5}\;M_\star$, which falls at the frontier
between regime (ii) and (iii). The disc has $\alpha=3/2$ and
$\beta=1.24$, so that $x_p/\lambda_c\approx 1.38$. For this parameter,
we expect, from Fig.~\ref{f:xp}, that the normalized heating torque
for a low mass planet is about $0.45$. We recover this result in
Fig.~\ref{fig:tqhm}. The normalized torque for the other planets have
a smaller value. In addition, for these masses, large fluctuations
appear a few orbits after the insertion of the planet, and almost
immediately for the largest mass, due to the appearance of vortices on
the edge of the horseshoe region. As the heating torque plateaus in
less than one orbital period, we use the temporal window (materialised
by a grey band in Fig.~\ref{fig:tqhm}) from $1.5$ to $5$~orbits to
obtain time averaged values, except for the largest mass where the
early onset of large fluctuations precludes any time averaging. We
find a normalized value of $0.27$ for the planet of mass
$10^{-6}\;M_\star$, i.e. about $60$~\% of the low-mass value, and a
normalized value of $0.11$ for the planet of mass
$3\cdot 10^{-6}\;M_\star$ (i.e. an Earth mass planet if the central
star has a solar mass), which represents $25$~\% of the low-mass
value. Fig.~\ref{fig:lbhm} shows the density perturbation arising from
heat release for these three planets, with same values for the
isocontours. Comparison between the low-mass case and the second case
(with $M_p=10^{-6}\;M_\star$) shows that the isocontours are less
extended in this second case: with same luminosity, this more massive
planet heats less its surroundings than the low-mass planet, which is
in agreement with the smaller normalized torque. Although there is
some difference in the shape of the isocontours of these two cases, we
believe that the torque reduction for this second case arises
primarily from the fact that $M_p>M_c$, in much the same way as the
heating force is cut off for when $M_p\gtrsim M_c$ in studies of
(negative) dynamical friction \citet{2020arXiv200413422V}: when
$M>M_c$, not all the energy deposited in the gas by the planet ends up
as an excess of internal energy in a heated region near the perturber
(a hot trail in that work, a two-lobe pattern here). In the
third case, both trends are confirmed: the isocontours are even less
extended, and their distortion with respect to the other cases is now
evident. Studying the torque decay for masses in excess of $M_c$ is
therefore a twofold problem, which requires an assessment of the
decrease of the heating efficiency of the nearby gas and of the
distortion of the thermal lobes by the horseshoe flow. Finally, for
the largest mass considered, we do not find any lobed structure. A
localised, hot region is found, associated to a vortex that drifts in
the horseshoe region. There is no sizeable heating torque, even one
orbital period after the insertion of the planet in the disc.

\section{Discussion}
\label{sec:discussion}
\subsection{Relevance of thermal torques in different contexts}
\label{sec:relev-therm-torq}
There has been two recent attempts to include and assess the impact of
thermal torques in models of planetary population
synthesis. \citet{2019MNRAS.486.5690G} find that the heating torque
has a dramatic impact on scenarios of planetary migration and
formation, as low mass embryos undergo a sustained phase of outward
migration before superseding the critical mass, at which point they
reverse their migration toward the central
object. \citet{2020arXiv200400874B} perform an analysis with
prescriptions for the thermal torques similar to those of
\citet{2019MNRAS.486.5690G}, and find them to have a negligible impact
on the resulting planetary populations.  \citet{2020arXiv200400874B}
claim that the different outcome between their work and that of
\citet{2019MNRAS.486.5690G} stems from the use, in the latter, of a
large, constant pebble accretion rate of
$10^{-5}\;\mathrm{M}_\oplus.\mathrm{yr}^{-1}$, thereby overestimating
the luminosity of the embryos. This claim, however, is not correct, as
such a large accretion rate was used only to produce Fig.~6 of that
paper, while elsewhere the accretion rate was determined using
standard prescriptions for pebble accretion
\citep{2014A&A...572A..35L}.  It is therefore unclear where the
difference between apparently similar setups comes from. We stress,
however, that both works assume that thermal torques vanish abruptly
as soon as the planetary mass supersedes the critical mass of
Eq.~\eqref{eq:11} (or even this critical mass divided by $\sqrt\gamma$
in the case of \citet{2020arXiv200400874B}, who use the isothermal
sound speed to estimate $M_c$). This is a very stringent assumption,
at odds with the findings of
section~\ref{sec:toward-larger-masses}. While this can be regarded as
a conservative assumption in the case of \citet{2019MNRAS.486.5690G},
as it goes against their conclusion, such is not the case for
\citet{2020arXiv200400874B}.  Consider the run with an Earth-mass
planet of section~\ref{sec:some-illustr-runs}. Fig.~\ref{fig:tqhm}
shows that a heating torque of magnitude $\sim 2\Gamma_0h$ is exerted
on the planet if it has the critical luminosity $L_c$, so that the net
torque changes sign\footnote{We assume here that the Lindblad plus
  corotation torques amount to $-2\Gamma_0h/\gamma$.} for a luminosity
$L\sim (1+\gamma^{-1})L_c$. Specializing to the case of a solar mass
central star and an orbital radius of $r_p=5.2$~au, the data used in
our runs translates to a critical luminosity as low as
$L_c=7.8\times 10^{25}$~erg.s$^{-1}$. Making the simplifying
assumption that the planetary luminosity is given by
\begin{equation}
  \label{eq:12}
  L=\frac{GM_p\dot M_p}{R_p},
\end{equation}
the luminosity $\sim (1+\gamma^{-1})L_c$ required to reverse migration
corresponds to a mass doubling time  $M_p/\dot M_p\sim 8.9\cdot 10^5$~yrs, which
largely exceeds the mass doubling time for an Earth mass embryo
subjected to planetesimal or pebble accretion. In other words, a luminous
Earth mass embryo embedded in our fiducial disc, despite having a mass
five times larger than the critical mass (or six times larger than
the threshold considered by \citet{2020arXiv200400874B}), is still
subjected to a vigorous heating torque that drives an outward
migration at a rate substantially faster than that given by
usual torque formulae that neglect thermal torques.

The magnitude of thermal torques in the regime of large masses
($M_p>M_c$) still warrants further work, and will probably be best
tackled through high resolution numerical simulations, due to the need
to resolve the Bondi sphere, where the flow is highly non-linear, and
to the intricacies linked to the interaction with the horseshoe
flow. Nonetheless, it is clear that thermal torques play a far more
important role for forming planets than envisioned in the early work
mentioned above. We also note that thermal torques fluctuate for the
planet masses larger than critical. Further work is also warranted to
determine whether the time averaged value of thermal torques coincides
with that measured at early time, and how the magnitude of these
fluctuations depend on the disc's parameter, in particular the
viscosity. \citet{2019A&A...626A.109C}. Note that complex migratory
behaviours, arising from a fluctuating heating torque on exerted on
super-Earths, have also been found by \citet{2019A&A...626A.109C} when
the opacity of the disc is not constant.

\subsection{On the relevance of thermal torques in AGN discs}
\label{sec:relev-therm-torq-1}
We now turn to a more speculative discussion about the migration of
stellar or intermediate-mass objects in the discs surrounding Active Galactic
Nuclei (AGN). The considerations outlined below are speculative mostly
because we apply the formulae for thermal torques, which have been
derived in discs where the radiation pressure is negligible, to media
dominated by radiation pressure \citep{2003MNRAS.341..501S}.
Note that previous work on migration in AGN discs has also used
classical torque formulae at Lindblad and corotation resonances,
derived for equations of state that are not that of radiation
pressure dominated flows \citep[e.g.][]{2019ApJ...878...85S}.

The importance of heating torques for the migration of massive stars
or accreting compact objects in the accretion discs has been discussed
by \citet{2020arXiv200503785H}. Here we give a few simple
relationships\footnote{Despite the different nature of the objects
  considered in this part, we stick to the notation introduced in
  section~\ref{sec:initial-conditions}, so that $M_\star$ is the mass of
  the central object and $M_p$ that of one of its satellites.} which
help assess quickly the importance of thermal torques (be it the
heating torque or the cold thermal torque) in the different regions of
these discs. We base our discussion on the fiducial model of
\citet[][namely the model presented in Fig.~2 of that
work]{2003MNRAS.341..501S}.

\begin{figure*}
  \centering
  \includegraphics[width=\columnwidth]{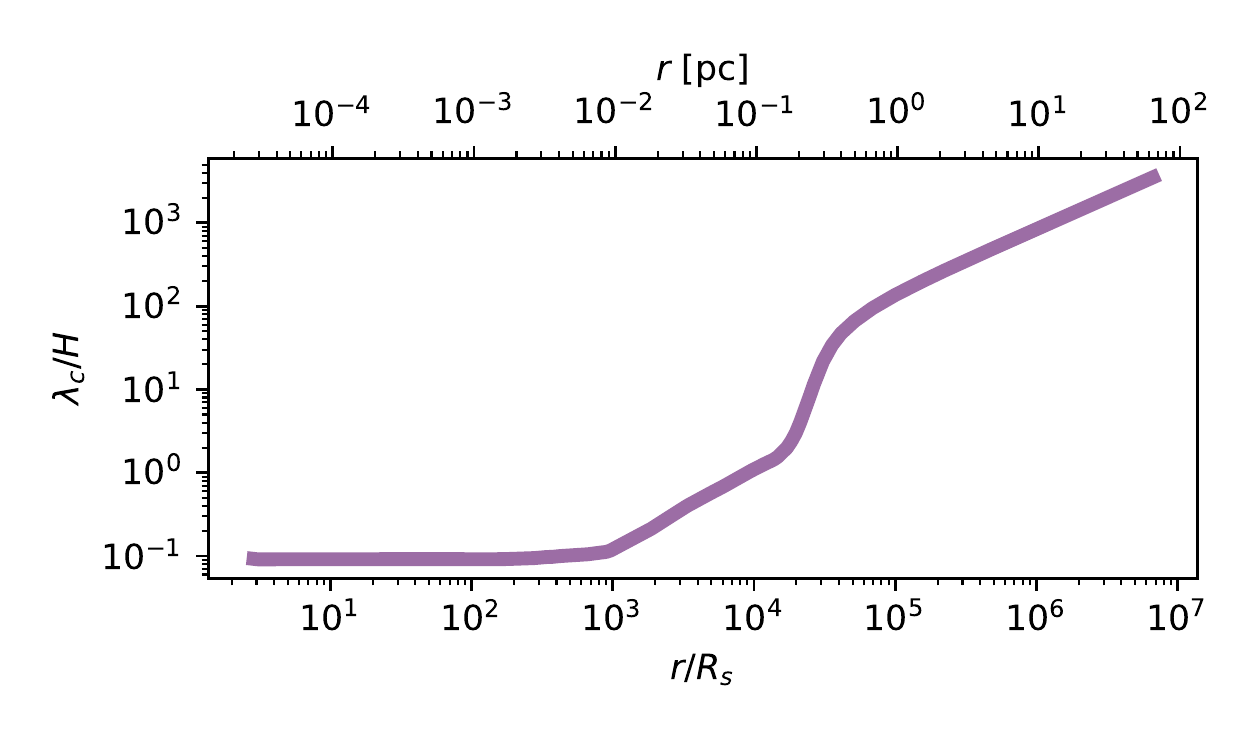}
  \includegraphics[width=\columnwidth]{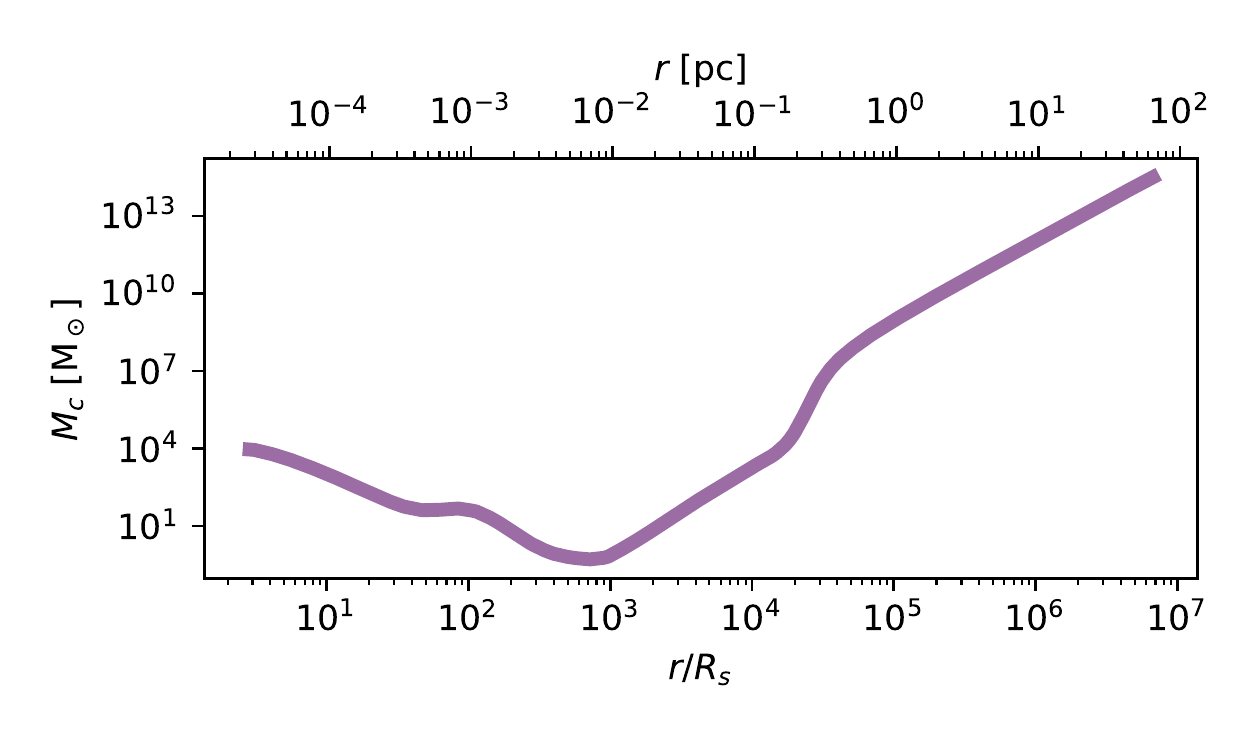}
  \caption{Left: ratio of the thermal to pressure lengthscale as a
    function of radial distance in the fiducial disc of
    \citet{2003MNRAS.341..501S}. This ratio needs to be small for
    thermal torques to be dominant. Right: critical mass as a function
    of radial distance in this disc. Thermal torques are cut-off for
    masses above the critical mass, as discussed in
    section~\ref{sec:toward-larger-masses}.}
  \label{fig:agn}
\end{figure*}

Thermal torques dominate over Lindblad and corotation torques when the
ratio of the pressure lengthscale $H$ to the size of the disturbance
$\lambda_c$ is large on the one hand, and when the mass of the
perturber is not large compared to the critical mass given by
Eq.~\eqref{eq:11}. Both quantities (the size of the disturbance and
the critical mass) depend on the thermal diffusivity. Over the whole
radial range considered by \citet{2003MNRAS.341..501S}, the total
pressure is dominated by the radiation pressure, hence the thermal
diffusivity has order of magnitude
\begin{equation}
  \label{eq:18}
  \chi \sim \bar\ell c,
\end{equation}
where $\bar\ell$ is the photon's mean free path and $c$ the speed of
light. Writing $\bar\ell\sim H/\tau$, where $\tau$ is the optical depth, and
using Eq.~\eqref{eq:lambda}, we are led to:
\begin{equation}
  \label{eq:19}
  \frac{\lambda_c}{H}\sim \frac{1}{\sqrt{h\tau}}\left(\frac{r}{R_s}\right)^{1/4},
\end{equation}
where $R_s$ is Schwarzschild's radius. We plot this ratio on the left
part of Fig.~\ref{fig:agn}. We see that $\lambda_c$ is smaller than
the disc's thickness up to $\gtrsim 10^4$ Schwarzschild's radii, or
approximately one tenth of a parsec. Up to this radius, thermal
torques may therefore supersede Lindblad and corotation torques,
whereas they should be negligible beyond that distance.

Next we turn to an estimate of the critical mass. Using
Eqs.~\eqref{eq:11} and~\eqref{eq:18}, we can write:
\begin{equation}
  \label{eq:20}
  M_c=\frac{\chi c_s}{G}=\frac{H c c_s}{\tau
    G}=\frac{c_s^3}{G\Omega\tau}\frac{c}{c_s}
  =M_\mathrm{th}\frac{1}{h\tau}\left(\frac{r}{R_s}\right)^{1/2}=M_\star \frac{h^2}{\tau}\left(\frac{r}{R_s}\right)^{1/2}
\end{equation}
We plot this mass on the right side of Fig.~\ref{fig:agn}. Over the
domain where we have seen that $\lambda_c < H$ (i.e. for
$r \lesssim 10^4R_s$), this critical mass is comprised between a few
solar masses and $10^4\;M_\odot$, the minimum being reached at $r\sim
10^3R_s\sim 10^{-2}$~pc.
This indicates that thermal torques should be important for stellar
mass objects within $0.1$~pc of the central object, while they should be
important for intermediate-mass black holes only on a fraction of this
domain. Whether thermal torques induce inward or outward migration
depends on which of the cold or heating torque dominates,
i.e. whether the object's luminosity is sub- or super-critical. Using
Eqs.~\eqref{eq:luminosity} and~\eqref{eq:18}, we can
express the critical luminosity as:
\begin{equation}
  \label{eq:21}
  L_c=\frac{4\pi GM_p\bar\ell c\rho_0}{\gamma}\sim \frac{4\pi GM_pc}{\kappa},
\end{equation}
where $\kappa\sim 1/(\ell\rho_0)$ is the opacity. The right hand side
of Eq.~\eqref{eq:21} is the Eddington luminosity of the perturber.
The critical luminosity is therefore of the order of Eddington's
luminosity. Objects radiating at luminosities comparable to
Eddington's could be subjected to a large heating torque and possibly
to a net, positive thermal torque, whereas for less luminous objects
the net thermal torque would be negative, inducing inward migration.

Some cautionary remarks are in order:
\begin{itemize}
\item Given the large variations of $h$ and $\tau$, markedly different
  results are to be expected if the model's parameters are
  changed. The above discussion is highly specific to the canonical
  model of \citet{2003MNRAS.341..501S}.
\item We reiterate that the analytic expressions for thermal
  torques have been obtained for a medium in which the radiation
  pressure is negligible.  Further studies are warranted to assess
  whether they take a similar expression in discs dominated by
  radiation pressure.
\item In the formulation of thermal forces in protoplanetary discs,
  the momentum injected into the medium by the luminous object is
  neglected, as the critical luminosity is order of magnitudes smaller
  than Eddington's \citep{2020arXiv200413422V}. Such is not the case
  here, and the injection of momentum should be taken into account in
  a formulation specific to AGN discs.
\end{itemize}

\section{Conclusions}
\label{sec:conclusions}
We have performed numerical simulations of low-mass planets embedded
in discs with thermal diffusion, aimed at checking analytical formulae
of thermal torques. We have confirmed that thermal torques depend
essentially on the distance $x_p$ between the planet and corotation.
We have also found a satisfactory agreement between analytics and
numerics when the distance of the planet to its corotation is much
smaller than the characteristic size of the thermal disturbance, and
we have confirmed that when the corotation offset becomes comparable
to the disturbance's size, thermal torques tend to plateau toward a
value given by a dynamical friction calculation. We have performed a
convergence study and found that the thermal disturbance must be
resolved over typically $10$~zones for an agreement at the $20-30$~\%
level between analytics and numerics. We have found a relatively large
discrepancy ($30$~\%) for the low value of $x_p$ that we used in the
convergence study. The expected torque value is then accordingly
small, which compounds any minute additional dependence of the thermal
torque that has been neglected in the analytic study of
\citet{2017MNRAS.472.4204M}, such as a dependence on the temperature
gradient, on the surface density gradient, or curvature effects.
The somehow relatively large residual found in that case may signal
such additional, small contributions to the heating torque not captured in
the analytic study, or simply that even higher resolutions than the
ones used here are required for a better agreement.

We have also evaluated the cold thermal torque, formerly called the
cold finger effect \citep{2014MNRAS.440..683L}, and found it to be in
good agreement with analytic estimates. We have checked that the
luminosity at which the net thermal torque switches from negative to
positive is within a few percents of the critical luminosity given by
Eq.~\eqref{eq:luminosity}.

We have also studied thermal torques for larger planet masses, that
supersede the critical mass $M_c$ (Eq.~\ref{eq:11}). We find a decay
of the torque, arising from a lower heating efficiency of the nearby
gas. Furthermore, we also find that as the planetary mass increases,
the horseshoe flow distorts the heated lobes. This can have an impact
on the value of the thermal torque, but we have not quantified this
effect. Overall, we find that thermal torques become insignificant for
planetary masses one order of magnitude larger than the critical
mass.

From our analysis we conclude that thermal torques can be largely
dominant over the Lindblad and corotation torques for objects in the
Earth mass range.  They are much more relevant for scenarios of planet
migration and formation than considered in previous work where they
were artificially and abruptly cancelled when the planetary mass
becomes larger than $M_c$. A systematic and quantitative assessment of
the torque decay in the large mass regime requires very high
resolution calculations, that resolve the Bondi radius of embedded
planets. These are beyond the scope of the present work.

\smallskip

It should also be kept in mind that objects with a luminosity larger
than the critical luminosity of Eq.~\eqref{eq:luminosity} are not only
subjected to a positive thermal torque but they also experience a
growth of eccentricity and inclination \citep{2017arXiv170401931E,2019MNRAS.485.5035F}. The migration path of such objects cannot be reduced to a simple outward motion given by the torque's value in this regime, and its characterisation requires significant further work.

We finally apply our findings to stellar or intermediate-mass objects
embedded in discs around AGNs. This exercise should be taken with a
pinch of salt, however, as these discs are usually dominated by
radiation pressure rather than gas pressure. Neither usual torque
formulae at Lindblad or corotation resonances, nor the formulae for
thermal torques have been derived for radiation dominated flows.

\section*{Acknowledgements}

R.O.C. acknowledges a postdoctoral CONACyT grant. Computational
resources were available thanks to a Marcos Moshinsky Chair and to
UNAM's PAPIIT grant BG101620. The authors thank the anonymous referee
for a thorough report which significantly improved the quality of the
manuscript.

\section*{Data Availability}
The data underlying this article will be shared on reasonable request to the corresponding author.


\bibliographystyle{mnras}


%
%


\bsp	
\label{lastpage}
\end{document}